\documentclass[twocolumn]{aastex61}
\usepackage{natbib}
\usepackage{multirow}
\usepackage{color}
\usepackage{bm}
\usepackage{soul}

\usepackage[normalem]{ulem}
\usepackage{amsmath}
\usepackage{amssymb}

\newcommand\aastex{AAS\TeX}

\received{\today}
\revised{}
\accepted{}
\submitjournal{\apjs}

\shorttitle{\aastex\ Energy level structure and transition data of Er$^{2+}$}
\shortauthors{Gaigalas et al.}

\begin{document}

\title{Energy level structure and transition data of E\textbf{\MakeLowercase{r}}$^{2+}$}

\correspondingauthor{Gediminas Gaigalas}
\email{gediminas.gaigalas@tfai.vu.lt}
\correspondingauthor{Pavel Rynkun}
\email{pavel.rynkun@tfai.vu.lt}

\author{Gediminas Gaigalas}
\affil{Institute of Theoretical Physics and Astronomy, 
Vilnius University, Saul\.{e}tekio Ave. 3, Lithuania}

\author{Pavel Rynkun}
\affil{Institute of Theoretical Physics and Astronomy, 
Vilnius University, Saul\.{e}tekio Ave. 3, Lithuania}

\author{Laima Rad\v{z}i\={u}t\.{e}}
\affil{Institute of Theoretical Physics and Astronomy, 
Vilnius University, Saul\.{e}tekio Ave. 3, Lithuania}

\author{Daiji Kato}
\affil{National Institute for Fusion Science, 
322-6 Oroshi-cho, Toki 509-5292, Japan}
\affil{Department of Advanced Energy Engineering, 
Kyushu University, Kasuga, Fukuoka 816-8580, Japan}

\author{Masaomi Tanaka}
\affil{Astronomical Institute, Tohoku University, Sendai 980-8578, Japan}

\author{P. J\"onsson}
\affil{Group for Materials Science and Applied Mathematics, Malm\"o University, SE-20506, Malm\"o, Sweden}



\begin{abstract}
The main aim of this paper is to present accurate energy levels of the ground [Xe]$4f^{12}$ and first excited [Xe]$4f^{11}5d$ configurations 
of Er$^{2+}$. 
The energy level structure of the Er$^{2+}$ ion was computed using the multiconfiguration Dirac-Hartree-Fock and relativistic configuration interaction (RCI) methods, as implemented in the GRASP2018 program package.
The Breit interaction, self-energy and vacuum polarization corrections were included in the RCI computations. The zero-first-order approach 
was used in the computations.
Energy levels with the identification in $LS$ coupling for all (399) states belonging to the [Xe]$4f^{12}$ and [Xe]$4f^{11}5d$ configurations are presented. Electric dipole (E1) transition data between the levels of these two configurations are computed. The accuracy of the these data are evaluated by studying the behaviour of the transition rates as functions of the gauge parameter as well as by evaluating the cancellation factors. The core electron correlations were studied using different strategies. 
Root-mean-square deviations obtained in this study for states of the ground and excited configurations from the available experimental or semi-empirical data are 649 cm$^{-1}$, and 747 cm$^{-1}$, respectively.
\end{abstract}



\section{Introduction}
Erbium is a lanthanide element with $Z=68$ and it has 6 stable isotopes. 
The isotopes are generated by different processes. Isotopes with $A=162$  
are produced by the $p$ process (proton capture), with $A=167,170$ by the $r$ process (rapid neutron capture),
with $A=164$ by the $p$ or the $s$ process (slow neutron capture) and with $A=166,168$ by the $r$ or the $s$
process \citep{JaschekJaschek1995}.
Since Er can be generated by the $r$ process, 
which can occur in the mergers of neutron star (NS), 
the atomic spectra of this element is of interest to a wide community of astrophysicists dealing with stellar nuclear synthesis.
The contribution of this element to the opacity of NS ejecta should be tested 
(e.g., \citealt{kasen17,tanaka18,tanaka19}), 
but even the energy levels of first excited configuration have not been fully presented.

Ions of erbium have been observed in different types of
stars. In the chemically peculiar (CP) stars, 
high abundances of lanthanide elements compared with solar values are observed.
In particular, Er III has been identified in the spectra of CP
stars of the upper main sequence (in the silicon star HD 192913 by \citet{Cowley_Crosswhite_1978}; 
 in the CP A star HR 465 by \citet{Cowley_Greenberg_1987}).  
\citet{Cowley_Mathys_1998} have identified lines in the range 5445-6587 \AA~in 
spectra of the extreme peculiar star HD 101065
 (Przybylski's star). In such stars the strongest spectral lines 
belong to the lanthanides rather than the iron 
group elements. In the above spectral range lines of Er III at $\lambda$ 6393.69, 5881.76 and 5988.39 \AA~appear.

The critical compilation of the energy levels of this ion, from \citep{Martin}, is based on a previous analysis by \citet{Spector}
of 24 levels for odd and 18 levels for even configurations, respectively.
Re-evaluation of the energy levels was done by  
\citet{Wyart_Blaise_Camus_1974a,Wyart_Koot_Van_Kleef_1974b,Wyart_Bauche-Arnoult_1981}.
For these investigations they used a semi-empirical parametric method.
More recently, the analysis of the spectrum of Er III was revised by 
\citet{Wyart}, and the number of identified energy levels increased from 45
to 115, including some levels of the $4f^{11}7s$ configuration.

\citet{Biemont} have measured radiative lifetimes of seven excited states of the $4f^{11}6p$ configuration using time-resolved
laser-induced fluorescence following two-photon excitation. Theoretical computation was done in frame of 
relativistic Hartree-Fock including core-polarization effects.

The aim of this paper is to provide accurate calculations of Er III, which can contribute to the stellar spectroscopy and understanding of opacities in NS mergers.
All levels of the ground [Xe]$4f^{12}$ and first excited [Xe]$4f^{11}5d$ configurations of Er$^{2+}$ are analysed in this paper.
Different core correlation effects and their inclusion strategies are presented.
The energy levels of these configurations and the corresponding electric dipole (E1) transition parameters were computed using
the GRASP2018 \citep{grasp2018} package. Computations are based on the multiconfiguration Dirac-Hartree-Fock (MCDHF) and relativistic configuration interaction (RCI) methods. 
The zero-first-order method was tested for various cases.

\section{General theory}

\subsection{Computational procedure}
The MCDHF method used in the present paper is
based on the Dirac-Coulomb (DC) Hamiltonian
\begin{eqnarray}
H_{DC} = \sum_{i=1}^N \left( c \; { \bm{\alpha} }_i \cdot
                    {\bf{ p }}_i
         + (\beta_i -1)c^2 + V^N_i \right)
         + \sum_{i>j}^N \frac{1}{r_{ij}},
\end{eqnarray}
where $V^N$ is the monopole part of the electron-nucleus Coulomb interaction, $\bm{ \alpha }$ and $\beta$ are the $4 \times 4$ Dirac matrices, and $c$ is the speed of light in atomic units.
The atomic state functions (ASFs) were obtained as linear
combinations of symmetry adapted configuration state functions (CSFs)
\begin{equation}
\label{ASF}
\Psi({\mathit \gamma} PJM)  = \sum_{j=1}^{N_{CSFs}} c_{j} \Phi(\gamma_{j}PJM).
\end{equation}
Here $J$ and $M$ are the angular quantum numbers and $P$ is parity.
$\gamma_j$ denotes other appropriate labeling of the configuration state function $j$,
for example orbital occupancy and coupling scheme. Normally the label ${\mathit \gamma}$ of the atomic state
function is the same as the label of the dominating CSF, see also section \ref{RCI}.
For these calculations the spin-angular approach \citep{Gaigalas_1996,Gaigalas_1997},
which is based on the second quantization in coupled
tensorial form, on the angular momentum theory in
three spaces (orbital, spin, and quasispin) and on the reduced coefficients of fractional parentage,
was used. It allows us to study configurations with open $f$-shells without any restrictions.
The  CSFs are built from products of one-electron Dirac orbitals. 
Based on a weighted energy average of several states, the so-called extended optimal level (EOL) scheme \citep{EOL},
both the radial parts of the Dirac orbitals and the 
expansion coefficients were optimized to self-consistency in the relativistic self-consistent field procedure \citep{topical_rev}.

\subsection{Zero-first-order method}

The CSF space can be divided into two parts according to Brillouin-Wigner perturbation theory \citep{Lindgren,Kato_2001}:

i) a principal part ($P$), which contains CSFs that account for the major parts of the wave functions and is referred to as a zero-order partitioning; 

ii) an orthogonal complementary part ($Q$), which contains CSFs that represent minor corrections and 
is referred to as a first-order partitioning.

%
Interaction between $P$ and $Q$ is assumed to be the lowest-order
perturbation. The total energy functional is partitioned into the
zero-order part ($H^{(0)}$) and the residual part ($V$). The Dirac-Fock 
energy functional is chosen as the zero-order part; the
residual part then represents a correlation energy functional.
The second-order Brillouin-Wigner perturbation theory then
leads to,
\begin{eqnarray}
(E-H^{(0)}_{QQ})^{-1}V_{QP}\Psi_P = \Psi_Q,
\nonumber \\
\left[H^{(0)}_{PP} + V_{PP} + V_{PQ}(E-H^{(0)}_{QQ})^{-1}V_{QP}\right]\Psi_P = E \Psi_P .
\end{eqnarray}
The above equations define the first-order correlation operator 
and the second-order effective Hamiltonian operator for
the {$P$-space}, respectively. In the brackets of the second equation, 
the first and second terms compose the total energy
functional in the {$P$-space}, and the third term represents the
second-order correction to the correlation energy functional
in the $P$-space. The non-linear effective Hamiltonian equation 
is written in a linearized form,
\begin{equation}
\label{Hamiltonian}
\begin{pmatrix} H^{(0)}_{PP} +V_{PP} ~~~~V_{PQ} \\ V_{QP}~~~~~~~~~~~~~~~H^{(0)}_{QQ}\end{pmatrix} 
\begin{pmatrix} \Psi_P \\ \Psi_Q \end{pmatrix}
=
E\begin{pmatrix} \Psi_P \\ \Psi_Q \end{pmatrix} .
\end{equation}
The requirement that the total energy functional ($E$) is
stationary with respect to variations in spin-orbitals ($\{\phi\}$) under 
the normalization and the orthogonality conditions leads
to a set of the Euler-Lagrange equations,
\begin{eqnarray}
\frac{\delta E [\{\phi\}]}{\delta \phi_a} = \mu_a \phi_a + \sum_{b \neq a} \mu_{ab}\phi_b \;,
\end{eqnarray}
where $\{\mu\}$ are the Lagrange multipliers. The above equations
are nothing but reduced MCDHF equations. That is to say, an
apparent connection between the second-order Brillouin-Wigner
perturbation energy functional and a set of reduced 
MCDHF equations is provided.

Block $H_{QQ}^{(0)}$ is diagonal in the Hamiltonian matrix (eq. \ref{Hamiltonian}).
As a result, computation time and size required 
for the construction of the Hamiltonian matrix are reduced. 
This method, named as zero-first-method (ZF), has the potential for
taking a very large configuration space into account, 
which is almost unachievable by full MCDHF and RCI methods, 
and for allowing accurate calculation to be performed with relatively small computational resources 
provided the $Q$-space contributes perturbatively to 
the $P$-space.

\subsection{Relativistic configuration interaction method}
\label{RCI}
The RCI method taking
into account Breit and quantum electrodynamic (QED) corrections \citep{grant, topical_rev}, was used in the computations.
The transverse photon interaction (Breit interaction)
\begin{eqnarray}
\label{eq:Breit}
         H_{\mbox{{\footnotesize Breit}}} =  - \sum_{i<j}^N \left[ \bm{\alpha}_{i} \cdot \bm{\alpha}_{j}\frac{ \cos(\omega_{ij} r_{ij}/c)}{r_{ij}}  \right.
			\nonumber \\
         + \left. (\bm{\alpha}_{i} \cdot \bm{\nabla}_i ) (\bm{\alpha}_{j} \cdot \bm{\nabla}_j )\frac{ \cos(\omega_{ij}r_{ij}/c) -1}{\omega_{ij}^2 r_{ij}/c^2} \right]
\end{eqnarray}
was included in the Hamiltonian. The photon frequencies $\omega_{ij}$, used for calculating
the matrix elements of the transverse photon interaction,
were taken as the difference of the diagonal Lagrange multipliers
associated with the Dirac orbitals \citep{Breit}. 

In the present calculations, the ASFs were obtained as expansions
over $jj$-coupled CSFs. To provide the $LSJ$ labeling system,
the ASFs were transformed from a $jj$-coupled CSF basis into an $LSJ$-coupled CSF basis using the
method developed by \citet{jj2lsj_atoms}.

\subsection{Computation of transition parameters}
The evaluation of radiative electric dipole (E1) transition data (transition probabilities, oscillator
strengths) between two states: $\gamma' P'J'M'$ and $\gamma PJM$, built on different and independently optimized orbital sets is non-trivial. 
The transition data can be expressed in terms of the transition moment, which is defined as
\begin{eqnarray}
\langle \,\Psi(\gamma PJ)\, \|  {\bf T}^{(1)} \| \,\Psi(\gamma' P'J')\, \rangle  &=&
\nonumber \\
 \sum_{j,k} c_jc'_k \; \langle \,\Phi(\gamma_j PJ)\, \|  {\bf T}^{(1)} \| \,\Phi(\gamma'_k P'J')\, \rangle,
\end{eqnarray}
where ${\bf T}^{(1)}$ is the transition operator. 
The calculation of the transition moment breaks down to the task of 
summing up reduced matrix elements between different CSFs.
The reduced matrix elements can be evaluated using standard techniques assuming
that both left and right hand CSFs are formed from the
same orthonormal set of spin-orbitals. This constraint is severe, since a high-quality and compact
wave function requires orbitals optimized
for a specific electronic state, for an example, see \citep{SF}.
To get around the problems of having a single orthonormal set of spin-orbitals, the wave function representations
of the two states, i.e. $\gamma' P'J'M'$ and $\gamma PJM$ were transformed in such way that the orbital sets became biorthonormal \citep{biotra}.
Standard methods were then used to evaluate the matrix elements of the transformed CSFs.

The reduced matrix elements are expressed via spin-angular coefficients $d^{(1)}_{ab}$ and operator strengths as:
\begin{eqnarray}
\langle \,\Phi(\gamma_j PJ)\, \|  {\bf T}^{(1)} \| \,\Phi(\gamma'_k P'J')\, \rangle  &=&
\nonumber \\
 \sum_{a,b} d^{(1)}_{ab} \; \langle \,n_al_aj_a\, \|  {\bf T}^{(1)} \| \,n_bl_bj_b\, \rangle.
\end{eqnarray}
Allowing for the fact that we are now using Brink-and-Satchler type reduced matrix elements, we have
\begin{eqnarray}
\langle \,n_al_aj_a\, \|  {\bf T}^{(1)} \| \,n_bl_bj_b\, \rangle  &=&
\nonumber \\
\left( \frac{(2j_b + 1)\omega}{\pi c} \right) ^{1/2} (-1)^{j_a - 1/2}  \begin{pmatrix}j_a~~~~1~~~~j_b \\ \frac{1}{2}~~~~0~-\frac{1}{2}\end{pmatrix} \overline{M_{ab}} 
\end{eqnarray}
where $\overline{M_{ab}}$, is the radiative transition integral defined by \citet{gauge}. The latter integral can be written
$\overline{M_{ab}} = \overline{M^e_{ab}} + G\overline{M^l_{ab}}$, where
$G$ is the gauge parameter. When $G=0$ we get the Coulomb (velocity) gauge, whereas for $G=\sqrt{2}$ we get the Babushkin (length) gauge.
In the general case, the gauge dependence has a parabolic form with
respect to the gauge parameter ($G$ axis) \citep{Rudzikas,Gaigalas_2010_PRA}. This dependence
may also be used for the evaluation of the accuracy of the results.
The more accurate the wave functions, the closer the parabola is to a straight line.

For electric dipole transitions the Babushkin and Coulomb gauges give the same
value of the transition moment for exact solutions of the Dirac-equation \citep{gauge}.
For approximate solutions the transition moments differ, and the quantity $dT$, defined as \citep{Ekman_2014}
\begin{equation}
\label{accuracy}
dT = \frac{|A_l-A_v|}{\max(A_l,A_v)},
\end{equation}
where $A_l$ and $A_v$ are transition rates in length and velocity form, can be used as a measure of the uncertainty of the
computed rate.

In the present work also the cancellation factor (CF), which shows cancellation effects in 
the computation of transition parameters was investigated. The cancellation factor is defined as \citep{Cowan,CF_Zhang}
\begin{eqnarray}
CF =  \left(\frac{\left|\sum_{j}\sum_{k} c_j \; \langle \,\Phi(\gamma_j PJ)\, \|  {\bf T}^{(1)} \| \,\Phi(\gamma'_k P'J')\, \rangle c'_k\right|}
{\sum_{k} \sum_{j} \left|c_j \; \langle \,\Phi(\gamma_j PJ)\, \|  {\bf T}^{(1)} \| \,\Phi(\gamma'_k P'J')\, \rangle c'_k\right|} \right)^2 .
\end{eqnarray}
To calculate CFs some modifications to the GRASP2018 \citep{grasp2018} package were done.
A small value of the CF, for example less than 0.1 or 0.05 (values are given in \citep{Cowan}), indicates that the calculated transition parameter, such as transition rate or oscillator strength, is affected
by a strong cancellation effect. Transition parameters with small CF are often associated with large uncertainties.

\section{Computational strategies}
\label{comp_strategies}
The study of the Er$^{2+}$ ion, as well as of the other lantanides, is quite a complex task because of the open $f$ shells. For systems with open $f$ shells, the number of CSFs increases very rapidly when including various electron correlation effects.
Computations for such systems using standard schemes are extremely demanding.
For this reason new computational strategies were developed and tested for Er$^{2+}$.

To obtain good wave functions, various electron correlation effects were investigated.
The ZF method was applied to reduce computational resources in different steps of the calculations
and to facilitate the inclusion of more electron correlation effects.
The final wave functions were used to compute electric dipole (E1) transition data between the levels of the two configurations.
The computational strategies will be discussed in more details in the sections below.

\subsection{Generation of initial wave functions and active space construction}
\label{Swf}
The first step of the wave function generation was an MCDHF computation of the [Xe]$4f^{12}$ configuration. 
In the second step, orbitals from the first step were kept frozen and used for the [Xe]$4f^{11}5d$ configuration, for which 
only the $5d$ orbitals ($5d+$ and $5d-$ in relativistic notation) were optimized.
In the tables, such an initial  computation in two steps will be referred to as a computation for the multireference (MR) space of CSFs.
The orbitals belonging to the [Xe]$4f^{12}$ configuration were kept frozen to get correct order 
for the states of the ground and excited configurations. 
A similar technique for the generation of the initial wave functions was already applied for Nd ions \citep{Gaigalas_2019}.

In the following steps of the computation, active spaces (AS) of CSFs were generated by 
allowing single-double (SD) or single-restricted-double (SrD) substitutions from only the valence shells or from valence and core shells of both configurations
to the orbital spaces (OS): $OS_1 = \{6s,6p,6d,5f\}$, ..., $OS_4 = \{9s,9p,9d,8f,7g,7h\}$.
When a new OS is being computed, the previous orbitals are frozen.
In Table \ref{summary} the number of CSFs used in the computations for the even and odd states is given.
The strategies mentioned in this Table will be described below in greater detail.

The Breit interaction and QED effects were included in RCI calculations. These corrections were taken into account in all strategies.

\begin{deluxetable}{r r r c}[ht!!]
\tablecaption{\label{summary} Summary of Active Space Constructions for the MCDHF and RCI Computations.}
\tablehead{
&\multicolumn{2}{c}{No. of CSFs}   \\ \cline{2-3}
  \multicolumn{1}{c}{Strategy and $AS$}&\multicolumn{1}{c}{Even} & \multicolumn{1}{c}{Odd} &  \multicolumn{1}{c}{ZF} }
\startdata
\textbf{SD 4f} $AS_1$ &  25~618 &   407~606 &$P$ \\
               $AS_2$ & 115~146 & 2~414~665 &$P+Q$ \\
               $AS_3$ & 326~187 & 7~986~088 &$P+Q$ \\  
							 $AS_4$ & 649~673 &16~859~203 &$P+Q$ \\ \hline 
\textbf{SD 5d} $AS_1$ &  25~618 &   538~902 &$P$ \\
               $AS_2$ & 115~146 & 2~868~718 &$P+Q$ \\ 
               $AS_3$ & 326~187 & 8~958~563 &$P+Q$ \\
							 $AS_4$ & 649~673 &18~527~744 &$P+Q$ \\ \hline 
\textbf{SD 5p} $AS_2$ & 369~343 & 11~769~255 \\ \hline 
\textbf{SD 5s} $AS_2$ & 193~028 & 4~745~781 \\ \hline 
\textbf{SrD 5p 5d} $AS_2$ & 337~325 & 10~720~590 \\ \hline		
\textbf{SD 5s 5d} $AS_2$ & 193~028 & 5~584~829 \\ \hline		
\textbf{SrD 5s 5p 5d} $AS_2$ & 414~383 & 13~402~965 \\ \hline		
\textbf{SD 5s 5p 5d} $AS_2$ & 476~274 & 19~482~860 \\ 		
\hline
\enddata
\tablecomments{The number of CSFs for the even and odd parities are given for each computational strategy and AS.}
\end{deluxetable}

\subsection{Valence-valence electron correlations}
Two strategies for including valence-valence (VV) electron correlations were investigated. 
In the first, the \textbf{SD 4f} strategy, the orbitals of which were used in all other strategies 
(\textbf{SD 5d}, \textbf{SD 5p}, \textbf{SD 5s}, \textbf{SrD 5p 5d}, 
\textbf{SD 5s 5d}, \textbf{SrD 5s 5p 5d}, \textbf{SD 5s 5p 5d} for these only RCI computations were performed), 
SD substitutions were allowed only from
the $4f$ valence shell of both configurations to the different orbital spaces. Later, separate computations were done for $AS_2$ for the even and odd parities 
and continued for the $AS_{3}$, built from the $OS_3$ orbital space. In the second strategy, the \textbf{SD 5d} strategy, SD substitutions were allowed from both valence ($4f$ and $5d$) shells to the different orbital spaces. Results of these investigations are presented in Table \ref{wf} 
and will be discussed in section \ref{VV_results}.
\subsection{Core-valence and core-core electron correlations}
The contribution of core-valence (CV) and core-core (CC) electron correlation effects to the energy levels was studied in RCI calculations
by allowing SD or SrD substitutions from core ($5p, 5s$) shells.
Results of these computations are presented in Table \ref{wf2}.
The orbital spaces are the same as described in section \ref{Swf}.
The column labeling is similar, for example, the notation \textbf{SD 5p} means that SD substitutions were done from the $4f$ and $5p$ shells.
In some computational schemes restrictions for the substitutions were applied.  
SrD substitutions in the \textbf{SrD 5p 5d} strategy mean that 
SD substitutions were done from the $4f$ and $5d$ shells, but from the $5p$ 
shell only S substitutions were allowed. 
In the \textbf{SrD 5s 5p 5d} strategy restrictions are applied to the $5s$ and $5p$ shells by allowing only S substitutions from these shells.

A summary of the active spaces of the different strategies, including core-valence and core-core electron 
correlation, is displayed in Table \ref{summary}.
From the Table it is seen that substitutions from core shells rapidly increase the number of CSFs.
The contribution of these correlations effects to energy levels will be presented in Section \ref{CC_CV_results}.

\subsection{Electron correlations using the zero-first-order method}
\label{ZF_computation}
The ZF method was applied to the \textbf{SD 4f} and \textbf{SD 5d} strategies and tested at different steps of the computations to 
reduce the computational load.
These results are presented in Tables \ref{ZF_contribution} and \ref{ZF_contribution_core}.
Firstly, the ZF method was applied 
to the MCDHF calculation in the \textbf{SD 4f} strategy for $AS_2$.
The results of these calculations, performed separately for the  even and odd configurations, are marked as \textbf{ZF$^{MCDHF}$}.
For the $AS_{2,3,4}$ active spaces the $AS_1$ space was used as the principal ($P$) part.
The principal part was selected based on the convergence of the energies, see section \ref{VV_results}. 
The sizes of the $P$ and $P+Q$ spaces used in the calculations are given in Table \ref{summary}.
Orbitals from the \textbf{SD 4f ZF$^{MCDHF}$} strategy were used 
in the RCI calculations for the \textbf{SD 4f ZF$^{MCDHF}_{RCI}$}, \textbf{SD 5d ZF$^{MCDHF}$}, and \textbf{SD 5d ZF$^{MCDHF}_{RCI}$} strategies.

The ZF approach was also used in the RCI calculations.
The results are displayed in Tables \ref{ZF_contribution}, \ref{ZF_contribution_core} and referred to as \textbf{ZF$_{RCI}$}.
The last columns of the Tables present the results of
RCI computations using the ZF method
based on orbitals from the  \textbf{ZF$^{MCDHF}$} calculations. These results 
are referred to as \textbf{ZF$^{MCDHF}_{RCI}$}. 

\section{Energy levels results}
\label{results}
Parts of the computed energy spectra from different strategies (described in section \ref{comp_strategies}) are presented in Tables \ref{wf} - \ref{energy_final}. 
The labels of the energy levels are given in $LS$ notation which are taken from NIST \citep{NIST}, or ordered by energy values for fixed $J$ value (POS). 
The notation $4f^N~^{(2S+1)} L ^{Nr} ~n'l' ~^{(2S'+1)}L'$ is used for the level labels. 
Intermediate quantum numbers define parents levels $4f^N~^{(2S+1)} L ^{Nr}$, 
where $N$ is electron number in the $4f$ shell, $(2S+1)$ is multiplicity, 
$Nr$ is a sequential index number representing the group labels $\nu$$WU$ for the term, 
and $L$ is orbital quantum number (see \citet{senioriy_2} in more details).
Energies in parentheses are from semi-empirical (SE) calculations by \citet{Wyart}. 
The total amount of energy levels presented in the NIST \citep{NIST} database and 
in the paper \citep{Wyart} for the ground and first excited configuration is only 64.
The accuracy of computed energy spectra was evaluated by comparing results with the NIST/(SE) data and calculating 
the relative difference $\Delta E/E =  (E_{NIST/(SE)} - E) / E_{NIST/(SE)} $.

\subsection{Convergence and valence-valence electron correlations}
\label{VV_results}
Table \ref{wf} displays the results when just VV correlations (\textbf{SD 4f} and \textbf{SD 5d} strategies) are included.
Using the \textbf{SD 4f} strategy we infer that the wave function relaxation for $AS_2$, resulting from separate computations for the even and odd parities, in comparison to the computations where
the even and odd parities are computed together, has small effect on the energy levels. 
It moderately increases the transition energy value by 0.15\% 
(0.09\% for levels of ground configuration and 0.15\% for levels of excited configuration).
For this comparison all 399 levels were included. 

\begin{deluxetable*}{lrlrrrrr@{\hskip 0.05pt}rr@{\hskip 0.05pt}rrr@{\hskip 0.05pt}rr@{\hskip 0.05pt}r}[ht!!]
\tabletypesize{\tiny}
\tablecaption{\label{wf} Energy Levels from RCI Calculations Using the \textbf{SD 4f} and \textbf{SD 5d} Strategies.}
\tablehead      {
$LS$& POS &\multicolumn{1}{c}{JP}&\multicolumn{1}{c}{NIST/(SE)}&&\multicolumn{6}{c}{\textbf{SD 4f}} && \multicolumn{4}{c}{\textbf{SD 5d}}\\
\cline{6-11} \cline{13-16} 
&&&&&\multicolumn{2}{c}{Orthogonal} \\
\cline{6-7}
&&&                                        & MR     &$AS_1$       &$AS_2$ &\multicolumn{2}{c}{$AS_2$} &\multicolumn{2}{c}{$AS_3$} &&\multicolumn{2}{c}{$AS_2$} &\multicolumn{2}{c}{$AS_3$}
}
\startdata
$4f^{12}~{}^3H$               & 1 & 6+&     0.00 &     0  & 0     &     0 & 0     &         &     0 &          && 0     &         &     0 \\          
$4f^{12}~{}^3F$               & 1 & 4+&  5081.79 &  6335  & 6142  &  5898 & 5895 /&$-$16.00 &  5744/&$-$13.03  && 5895 /&$-$16.00 & 5744 /&$-$13.03 \\
$4f^{12}~{}^3H$               & 1 & 5+&  6969.78 &  6673  & 6733  &  6786 & 6784 /&2.66     &  6805/&2.36      && 6784 /&2.66     &  6805/&2.36     \\
$4f^{12}~{}^3H$               & 2 & 4+& 10785.48 & 11089  & 11036 & 10957 & 10958/&$-$1.60  & 10889/&$-$0.96   && 10958/&$-$1.60  & 10889/&$-$0.96  \\
$4f^{12}~{}^3F$               & 1 & 3+&(12472.55)& 14166  & 13908 & 13557 & 13566/&$-$8.77  & 13282/&$-$6.49   && 13566/&$-$8.77  & 13282/&$-$6.49  \\
$4f^{12}~{}^3F$               & 1 & 2+&(13219.80)& 15578  & 15236 & 14815 & 14825/&$-$12.14 & 14446/&$-$9.28   && 14825/&$-$12.14 & 14446/&$-$9.28  \\
$4f^{12}~{}^1G$               & 3 & 4+&(18383.59)& 18387  & 18391 & 18377 & 18360/&0.13     & 18381/&0.01      && 18360/&0.13     & 18381/&0.01     \\
$4f^{11}~{}(^4I^1)~{}5d~{}^5G$& 1 & 6-& 16976.09 & 19978  & 15540 & 18872 & 18983/&$-$11.82 & 21480/&$-$26.53  && 14673/&13.56    & 16073/&5.32     \\
$4f^{11}~{}(^4I^1)~{}5d~{}^5H$& 1 & 7-& 17647.76 & 20984  & 16495 & 19770 & 19877/&$-$12.63 & 22337/&$-$26.57  && 15613/&11.53    & 17006/&3.64     \\
$4f^{11}~{}(^4I^1)~{}5d~{}^3L$& 1 & 9-& 18976.74 & 22664  & 17843 & 20827 & 20916/&$-$10.22 & 23382/&$-$23.21  && 16725/&11.87    & 18204/&4.07     \\
$4f^{11}~{}(^4I^1)~{}5d~{}^5I$& 1 & 8-& 19918.17 & 23811  & 19088 & 22181 & 22273/&$-$11.82 & 24699/&$-$24.00  && 17988/&9.69     & 19380/&2.70     \\
$4f^{11}~{}(^4I^1)~{}5d~{}^5L$& 1 &10-& 20470.13 & 23320  & 18548 & 21587 & 21673/&$-$5.88  & 24130/&$-$17.88  && 17644/&13.81    & 19158/&6.41     \\
$4f^{11}~{}(^4I^1)~{}5d~{}^5K$& 2 & 9-& 21688.17 &  25556 & 20625 & 23583 & 23664/&$-$9.11  & 26085/&$-$20.27  && 19561/&9.81     & 21006/&3.15     \\
$4f^{11}~{}(^4I^1)~{}5d~{}^5G$& 1 & 5-& 22016.77 &  25908 & 20848 & 24065 & 24164/&$-$9.75  & 26648/&$-$21.04  && 19890/&9.66     & 21285/&3.32     \\
$4f^{11}~{}(^4I^1)~{}5d~{}^5H$& 2 & 6-& 22606.07 &  26785 & 21696 & 24867 & 24963/&$-$10.43 & 27418/&$-$21.29  && 20716/&8.36     & 22102/&2.23     \\
$4f^{11}~{}(^4I^1)~{}5d~{}^5I$& 2 & 8-& 22951.42 &  27237 & 22101 & 25006 & 25085/&$-$9.30  & 27514/&$-$19.88  && 20903/&8.93     & 22321/&2.75     \\
$4f^{11}~{}(^4I^1)~{}5d~{}^5I$& 2 & 7-& 23302.78 &  81645 & 22982 & 25958 & 26039/&$-$11.74 & 28459/&$-$22.13  && 21733/&6.73     & 23103/&0.86     \\
$4f^{11}~{}(^4I^1)~{}5d~{}^5L$& 3 & 8-& 25482.12 &  30201 & 24501 & 27376 & 27453/&$-$7.73  & 29916/&$-$17.40  && 23276/&8.66     & 24757/&2.85     \\
$4f^{11}~{}(^4I^1)~{}5d~{}^5H$& 3 & 5-& 26192.66 &  31455 & 25931 & 29019 & 29109/&$-$11.13 & 31551/&$-$20.46  && 24808/&5.29     & 26162/&0.12     \\
$4f^{11}~{}(^4I^1)~{}5d~{}^5K$& 3 & 7-& 26579.91 &  31832 & 26197 & 29141 & 29220/&$-$9.93  & 31644/&$-$19.05  && 24959/&6.10     & 26367/&0.80     \\
                             &  1 & 4-&(26648.59)&  31020 & 25496 & 28690 & 28786/&$-$8.02  & 31252/&$-$17.27  && 24556/&7.85     & 25947/&2.63     \\
                             &  2 & 4-&(29469.40)&  34371 & 28921 & 32019 & 32112/&$-$8.97  & 34506/&$-$17.09  && 27877/&5.40     & 29198/&0.92     \\
                             &  3 & 4-&(30750.22)&  36329 & 31536 & 34481 & 34590/&$-$12.49 & 36669/&$-$19.25  && 30274/&1.55     & 31299/&$-$1.78  \\
                             &  4 & 4-&(32196.96)&  38241 & 32841 & 35729 & 35818/&$-$11.25 & 38077/&$-$18.27  && 31408/&2.45     & 32609/&$-$1.28  \\
                             &  5 & 4-&(33033.10)&  39637 & 33991 & 36713 & 36800/&$-$11.40 & 38990/&$-$18.03  && 32374/&2.00     & 33454/&$-$1.27  \\
                             &  6 & 4-&(35903.96)&  43048 & 37724 & 40326 & 40430/&$-$12.61 & 42276/&$-$17.75  && 36133/&$-$0.64  & 36946/&$-$2.90  \\
                             &  7 & 4-&(37608.12)&  45043 & 39093 & 41818 & 41896/&$-$11.40 & 44068/&$-$17.18  && 37438/&0.45     & 38533/&$-$2.46  \\
                             &  8 & 4-&(39667.36)&  46418 & 41110 & 43706 & 43809/&$-$10.44 & 45645/&$-$15.07  && 39556/&0.28     & 40364/&$-$1.76  \\
                             &  9 & 4-&(40580.40)&  46896 & 41790 & 44350 & 44448/&$-$9.53  & 46293/&$-$14.08  && 40165/&1.02     & 41000/&$-$1.03  \\
                             & 10 & 4-&(46937.23)&  47841 & 42249 & 44753 & 44848/&4.45     & 46678/&0.55      && 40498/&13.72    & 41266/&12.08    \\
                             &  3 & 9-&(27471.61)&  31838 & 25998 & 28861 & 28932/&$-$5.32  & 31400/&$-$14.30  && 24896/&9.37     & 26416/&3.84     \\
                             &  3 & 6-&(27472.46)&  32771 & 27231 & 30271 & 30355/&$-$10.49 & 34985/&$-$27.35  && 26082/&5.06     & 27443/&0.11     \\
                             &  4 & 6-&(28777.74)&  35231 & 29549 & 32540 & 32618/&$-$13.34 & 35063/&$-$21.84  && 28112/&2.31     & 29360/&$-$2.02  \\
                             &  5 & 6-&(30283.09)&  35487 & 29796 & 32595 & 32668/&$-$7.88  & 36114/&$-$19.25  && 28454/&6.04     & 29857/&1.41     \\
                             &  6 & 6-&(31095.82)&  36376 & 30755 & 33686 & 33767/&$-$8.59  & 36114/&$-$16.14  && 29549/&4.98     & 30894/&0.65     \\
                             &  7 & 6-&(33191.53)&  39104 & 34132 & 36805 & 36889/&$-$11.14 & 38955/&$-$17.36  && 32635/&1.68     & 33718/&$-$1.59  \\
                             &  8 & 6-&(33875.19)&  41285 & 35416 & 38012 & 38090/&$-$12.44 & 40207/&$-$18.69  && 33618/&0.76     & 34659/&$-$2.31  \\
                             &  9 & 6-&(35856.62)&  43309 & 37029 & 39631 & 39697/&$-$10.71 & 41952/&$-$17.00  && 35430/&1.19     & 36666/&$-$2.26  \\
                             & 10 & 6-&(36570.10)&  43513 & 38120 & 40683 & 40776/&$-$11.50 & 42673/&$-$16.69  && 36480/&0.25     & 37345/&$-$2.12  \\
                             &  3 & 5-&(27870.83)&  34210 & 28464 & 31164 & 31235/&$-$12.07 & 33646/&$-$20.72  && 26706/&4.18     & 27992/&$-$0.43  \\
                             &  4 & 5-&(29995.62)&  35316 & 29934 & 32954 & 33042/&$-$10.16 & 35388/&$-$17.98  && 28760/&4.12     & 30057/&$-$0.20  \\
                             &  5 & 5-&(31214.52)&  36805 & 31724 & 34665 & 34758/&$-$11.35 & 36967/&$-$18.43  && 30474/&2.37     & 31638/&$-$1.36  \\
                             &  6 & 5-&(32614.37)&  38071 & 33095 & 35892 & 35994/&$-$10.36 & 38025/&$-$16.59  && 31714/&2.76     & 32739/&$-$0.38  \\
                             &  7 & 5-&(33704.29)&  40276 & 34638 & 37366 & 37449/&$-$11.11 & 39616/&$-$17.54  && 33073/&1.87     & 34182/&$-$1.42  \\
                             &  8 & 5-&(36330.81)&  43191 & 37249 & 39952 & 40031/&$-$10.18 & 42203/&$-$16.16  && 35735/&1.64     & 36870/&$-$1.48  \\
                             &  9 & 5-&(36655.60)&  44613 & 38939 & 41535 & 41618/&$-$13.54 & 43638/&$-$19.05  && 37138/&$-$1.32  & 38054/&$-$3.81  \\
                             & 10 & 5-&(39265.81)&  47531 & 41949 & 44443 & 44536/&$-$13.42 & 46335/&$-$18.00  && 40220/&$-$2.43  & 41002/&$-$4.42  \\
                             & 11 & 5-&(40857.10)&  47534 & 42220 & 44961 & 45058/&$-$10.28 & 46878/&$-$14.74  && 40672/&0.45     & 41476/&$-$1.51  \\
                             & 12 & 5-&(46552.18)&  48691 & 42922 & 45419 & 45507/&2.25     & 47406/&$-$1.84   && 41210/&11.48    & 42034/&9.71     \\
                             & 13 & 5-&(48747.15)&  50473 & 44655 & 47228 & 47313/&2.94     & 49233/&$-$1.00   && 42975/&11.84    & 43871/&10.00    \\
                             &  4 & 8-&(28555.40)&  33718 & 27798 & 30612 & 30680/&$-$7.44  & 33109/&$-$15.95  && 26563/&6.98     & 28010/&1.91     \\
                             &  5 & 8-&(31701.46)&  36985 & 30873 & 33614 & 33681/&$-$6.24  & 36102/&$-$13.88  && 29611/&6.59     & 31074/&1.98     \\
                             &  4 & 7-&(28818.44)&  34225 & 28372 & 31253 & 31327/&$-$8.70  & 33744/&$-$17.09  && 27034/&6.19     & 28431/&1.34     \\
                             &  5 & 7-&(29610.99)&  35023 & 29079 & 31884 & 31953/&$-$7.91  & 34373/&$-$16.08  && 27752/&6.28     & 29159/&1.53     \\
                             &  6 & 7-&(32559.55)&  38684 & 32648 & 35309 & 35372/&$-$8.64  & 37743/&$-$15.92  && 31168/&4.27     & 32535/&0.08     \\
                             &  7 & 7-&(36636.87)&  40135 & 34251 & 36905 & 36973/&$-$0.92  & 39291/&$-$7.25   && 32864/&10.30    & 34212/&6.62     \\
                             &  1 & 3-&(29466.42)&  34203 & 28690 & 31847 & 31943/&$-$8.40  & 34353/&$-$16.58  && 27718/&5.93     & 29057/&1.39     \\
                             &  2 & 3-&(31846.16)&  37048 & 31225 & 34313 & 34404/&$-$8.03  & 36802/&$-$15.56  && 30216/&5.12     & 31554/&0.92     \\
                             &  3 & 3-&(33185.64)&  39693 & 34892 & 37685 & 37798/&$-$13.90 & 39691/&$-$19.61  && 33433/&$-$0.75  & 34279/&$-$3.29  \\
                             &  4 & 3-&(36167.30)&  43947 & 38081 & 40661 & 40736/&$-$12.63 & 42934/&$-$18.71  && 36191/&$-$0.07  & 37278/&$-$3.07  \\
                             &  5 & 3-&(37812.87)&  44279 & 39414 & 41981 & 42093/&$-$11.32 & 43938/&$-$16.20  && 37805/&0.02     & 38620/&$-$2.13  \\
                             &  6 & 3-&(38924.30)&  45611 & 40313 & 42866 & 42968/&$-$10.39 & 44779/&$-$15.04  && 38621/&0.78     & 39395/&$-$1.21  \\
                             &  7 & 3-&(40407.72)&  47120 & 41753 & 44221 & 44318/&$-$9.68  & 46177/&$-$14.28  && 39948/&1.14     & 40754/&$-$0.86  \\
                             &  1 & 2-&(38563.97)&  43843 & 31760 & 41645 & 34941/&9.39     & 37339/&3.17      && 30760/&20.24    & 32092/&16.78    \\	
\hline
\enddata 
\tablecomments{The relative difference compared with NIST/(SE) data is given in percent.}
\end{deluxetable*}

\begin{deluxetable*}{lrlrr@{\hskip 0.05pt}rr@{\hskip 0.05pt}rr@{\hskip 0.05pt}rr@{\hskip 0.05pt}rr@{\hskip 0.05pt}rr@{\hskip 0.05pt}rr@{\hskip 0.05pt}r}[ht!!]
\tabletypesize{\tiny}
\tablecaption{\label{wf2} Energy Levels from RCI Calculations Including CV and CC Electron Correlations.}
\tablehead      {
$LS$& POS &\multicolumn{1}{c}{JP}&\multicolumn{1}{c}{NIST/(SE)}
&\multicolumn{2}{c}{\textbf{SD 5p}}
&\multicolumn{2}{c}{\textbf{SD 5s}} 
&\multicolumn{2}{c}{\textbf{SrD 5p 5d}} 
&\multicolumn{2}{c}{\textbf{SD 5s 5d}} 
&\multicolumn{2}{c}{\textbf{SrD 5s 5p 5d}} 
&\multicolumn{2}{c}{\textbf{SD 5s 5p 5d}} \\
&&& & \multicolumn{2}{c}{$AS_2$}& \multicolumn{2}{c}{$AS_2$} & \multicolumn{2}{c}{$AS_2$}& \multicolumn{2}{c}{$AS_2$}&\multicolumn{2}{c}{$AS_2$}& \multicolumn{2}{c}{$AS_2$} 
}
\startdata
$4f^{12}~{}^3H$               & 1 & 6+&     0.00 & 0     &          & 0     &        & 0     &         & 0     &         & 0     &         & 0      \\ 
$4f^{12}~{}^3F$               & 1 & 4+&  5081.79 & 5816 /&$-$14.46  & 5816 /&$-$14.44& 5791 /&$-$13.96 & 5816 /&$-$14.44 & 5710 /&$-$12.37 & 5751 /&$-$13.16  \\
$4f^{12}~{}^3H$               & 1 & 5+&  6969.78 & 6773 /&2.83      & 6777 /&2.77    & 6784 /&2.67     & 6777 /&2.77     & 6778 /&2.75     & 6755 /&3.08      \\
$4f^{12}~{}^3H$               & 2 & 4+& 10785.48 & 10875/&$-$0.83   & 10894/&$-$1.01 & 10880/&$-$0.88  & 10894/&$-$1.01  & 10823/&$-$0.35  & 10809/&$-$0.22   \\
$4f^{12}~{}^3F$               & 1 & 3+&(12472.55)& 13387/&$-$7.33   & 13387/&$-$7.33 & 13362/&$-$7.13  & 13387/&$-$7.33  & 13184/&$-$5.70  & 13223/&$-$6.02   \\
$4f^{12}~{}^3F$               & 1 & 2+&(13219.80)& 14599/&$-$10.44  & 14603/&$-$10.47& 14576/&$-$10.26 & 14603/&$-$10.47 & 14354/&$-$8.58  & 14399/&$-$8.92   \\
$4f^{12}~{}^1G$               & 3 & 4+&(18383.59)& 18305/&0.43      & 18345/&0.21    & 18323/&0.33     & 18345/&0.21     & 18317/&0.36     & 18276/&0.58      \\
$4f^{11}~{}(^4I^1)~{}5d~{}^5G$& 1 & 6-& 16976.09 & 41934/&$-$147.02 & 19831/&$-$16.82& 14992/&11.69    & 14262/&15.99    & 14781/&12.93    & 18184/&$-$7.11   \\
$4f^{11}~{}(^4I^1)~{}5d~{}^5H$& 1 & 7-& 17647.76 & 42676/&$-$141.82 & 20691/&$-$17.24& 15783/&10.57    & 15143/&14.19    & 15518/&12.07    & 18929/&$-$7.26   \\
$4f^{11}~{}(^4I^1)~{}5d~{}^3L$& 1 & 9-& 18976.74 & 44039/&$-$132.07 & 21719/&$-$14.45& 17340/&8.63     & 16315/&14.03    & 17124/&9.76     & 20551/&$-$8.30   \\
$4f^{11}~{}(^4I^1)~{}5d~{}^5I$& 1 & 8-& 19918.17 & 45243/&$-$127.14 & 23062/&$-$15.78& 18460/&7.32     & 17476/&12.26    & 18151/&8.87     & 21547/&$-$8.18   \\
$4f^{11}~{}(^4I^1)~{}5d~{}^5L$& 1 &10-& 20470.13 & 44910/&$-$119.39 & 22487/&$-$9.85 & 18326/&10.47    & 17280/&15.59    & 18186/&11.16    & 21705/&$-$6.03   \\
$4f^{11}~{}(^4I^1)~{}5d~{}^5K$& 2 & 9-& 21688.17 & 46587/&$-$114.80 & 24423/&$-$12.61& 19982/&7.87     & 19003/&12.38    & 19648/&9.41     & 23133/&$-$6.66   \\
$4f^{11}~{}(^4I^1)~{}5d~{}^5G$& 1 & 5-& 22016.77 & 47034/&$-$113.63 & 24992/&$-$13.51& 20123/&8.60     & 19462/&11.60    & 19903/&9.60     & 23352/&$-$6.06   \\
$4f^{11}~{}(^4I^1)~{}5d~{}^5H$& 2 & 6-& 22606.07 & 47717/&$-$111.08 & 25765/&$-$13.97& 20849/&7.77     & 20236/&10.49    & 20578/&8.97     & 24017/&$-$6.24   \\
$4f^{11}~{}(^4I^1)~{}5d~{}^5I$& 2 & 8-& 22951.42 & 48071/&$-$109.45 & 25845/&$-$12.61& 21401/&6.75     & 20375/&11.23    & 21085/&8.13     & 24513/&$-$6.81   \\
$4f^{11}~{}(^4I^1)~{}5d~{}^5I$& 2 & 7-& 23302.78 & 48801/&$-$109.42 & 26809/&$-$15.05& 21988/&5.64     & 21201/&9.02     & 21667/&7.02     & 25102/&$-$7.72   \\
$4f^{11}~{}(^4I^1)~{}5d~{}^5L$& 3 & 8-& 25482.12 & 50537/&$-$98.32  & 28240/&$-$10.82& 23883/&6.28     & 22844/&10.35    & 23653/&7.18     & 27097/&$-$6.34   \\
$4f^{11}~{}(^4I^1)~{}5d~{}^5H$& 3 & 5-& 26192.66 & 51811/&$-$97.81  & 29902/&$-$14.16& 24893/&4.96     & 24334/&7.10     & 24629/&5.97     & 28094/&$-$7.26   \\
$4f^{11}~{}(^4I^1)~{}5d~{}^5K$& 3 & 7-& 26579.91 & 52063/&$-$95.87  & 30001/&$-$12.87& 25307/&4.79     & 24467/&7.95     & 25005/&5.92     & 28455/&$-$7.05   \\
                             &  1 & 4-&(26648.59)& 51645/&$-$93.80  & 29603/&$-$11.09& 24778/&7.02     & 24111/&9.52     & 24539/&7.92     & 27997/&$-$5.06   \\
                             &  2 & 4-&(29469.40)& 54881/&$-$86.23  & 32892/&$-$11.62& 28027/&4.89     & 27383/&7.08     & 27745/&5.85     & 31205/&$-$5.89   \\
                             &  3 & 4-&(30750.22)& 57119/&$-$85.75  & 35265/&$-$14.68& 30189/&1.83     & 29636/&3.62     & 29750/&3.25     & 33208/&$-$7.99   \\
                             &  4 & 4-&(32196.96)& 58399/&$-$81.38  & 36546/&$-$13.51& 31442/&2.34     & 30866/&4.14     & 31119/&3.35     & 34581/&$-$7.40   \\
                             &  5 & 4-&(33033.10)& 59209/&$-$79.24  & 37491/&$-$13.50& 32286/&2.26     & 31762/&3.85     & 31888/&3.47     & 35361/&$-$7.05   \\
                             &  6 & 4-&(35903.96)& 62783/&$-$74.86  & 41001/&$-$14.20& 35925/&$-$0.06  & 35384/&1.45     & 35404/&1.39     & 38898/&$-$8.34   \\
                             &  7 & 4-&(37608.12)& 64261/&$-$70.87  & 42577/&$-$13.21& 37261/&0.92     & 36805/&2.14     & 36852/&2.01     & 40347/&$-$7.28   \\
                             &  8 & 4-&(39667.36)& 66132/&$-$66.72  & 44349/&$-$11.80& 39308/&0.90     & 38758/&2.29     & 38730/&2.36     & 42249/&$-$6.51   \\
                             &  9 & 4-&(40580.40)& 66448/&$-$63.74  & 44935/&$-$10.73& 39656/&2.28     & 39219/&3.36     & 38922/&4.09     & 42477/&$-$4.67   \\
                             & 10 & 4-&(46937.23)& 67135/&$-$43.03  & 45403/&3.27    & 40249/&14.25    & 39676/&15.47    & 39660/&15.50    & 43170/&8.03      \\
                             &  3 & 9-&(27471.61)& 52079/&$-$89.57  & 29719/&$-$8.18 & 25527/&7.08     & 24483/&10.88    & 25331/&7.79     & 28847/&$-$5.01   \\
                             &  3 & 6-&(27472.46)& 53159/&$-$93.50  & 31131/&$-$13.32& 26362/&4.04     & 25550/&7.00     & 26031/&5.25     & 29456/&$-$7.22   \\
                             &  4 & 6-&(28777.74)& 54968/&$-$91.01  & 33363/&$-$15.93& 27916/&2.99     & 27508/&4.41     & 27532/&4.33     & 31022/&$-$7.80   \\
                             &  5 & 6-&(30283.09)& 55604/&$-$83.61  & 33434/&$-$10.40& 28912/&4.53     & 27990/&7.57     & 28635/&5.44     & 32078/&$-$5.93   \\
                             &  6 & 6-&(31095.82)& 56644/&$-$82.16  & 34524/&$-$11.03& 29950/&3.68     & 29034/&6.63     & 29659/&4.62     & 33103/&$-$6.45   \\
                             &  7 & 6-&(33191.53)& 59528/&$-$79.35  & 37535/&$-$13.08& 32780/&1.24     & 31945/&3.75     & 32302/&2.68     & 35782/&$-$7.80   \\
                             &  8 & 6-&(33875.19)& 60514/&$-$78.64  & 38733/&$-$14.34& 33550/&0.96     & 32915/&2.84     & 33090/&2.32     & 36572/&$-$7.96   \\
                             &  9 & 6-&(35856.62)& 62283/&$-$73.70  & 40367/&$-$12.58& 35579/&0.77     & 34759/&3.06     & 35118/&2.06     & 38603/&$-$7.66   \\
                             & 10 & 6-&(36570.10)& 63109/&$-$72.57  & 41345/&$-$13.06& 36358/&0.58     & 35683/&2.43     & 35777/&2.17     & 39291/&$-$7.44   \\
                             &  3 & 5-&(27870.83)& 53594/&$-$92.30  & 32008/&$-$14.84& 26576/&4.64     & 26237/&5.86     & 26324/&5.55     & 29827/&$-$7.02   \\
                             &  4 & 5-&(29995.62)& 55782/&$-$85.97  & 33802/&$-$12.69& 28973/&3.41     & 28212/&5.95     & 28626/&4.57     & 32048/&$-$6.84   \\
                             &  5 & 5-&(31214.52)& 57489/&$-$84.17  & 35477/&$-$13.66& 30691/&1.68     & 29870/&4.31     & 30284/&2.98     & 33712/&$-$8.00   \\
                             &  6 & 5-&(32614.37)& 58538/&$-$79.49  & 36642/&$-$12.35& 31708/&2.78     & 31056/&4.78     & 31280/&4.09     & 34760/&$-$6.58   \\
                             &  7 & 5-&(33704.29)& 59945/&$-$77.85  & 38120/&$-$13.10& 33071/&1.88     & 32417/&3.82     & 32637/&3.17     & 36108/&$-$7.13   \\
                             &  8 & 5-&(36330.81)& 62518/&$-$72.08  & 40690/&$-$12.00& 35739/&1.63     & 35088/&3.42     & 35311/&2.81     & 38797/&$-$6.79   \\
                             &  9 & 5-&(36655.60)& 63875/&$-$74.26  & 42241/&$-$15.24& 36869/&$-$0.58  & 36416/&0.65     & 36367/&0.79     & 39878/&$-$8.79   \\
                             & 10 & 5-&(39265.81)& 66853/&$-$70.26  & 45068/&$-$14.78& 40042/&$-$1.98  & 39334/&$-$0.17  & 39364/&$-$0.25  & 42872/&$-$9.18   \\
                             & 11 & 5-&(40857.10)& 67312/&$-$64.75  & 45623/&$-$11.66& 40402/&1.11     & 39864/&2.43     & 39789/&2.61     & 43274/&$-$5.92   \\
                             & 12 & 5-&(46552.18)& 67918/&$-$45.90  & 46074/&1.03    & 41104/&11.70    & 40375/&13.27    & 40491/&13.02    & 44015/&5.45      \\
                             & 13 & 5-&(48747.15)& 69613/&$-$42.80  & 47887/&1.76    & 42779/&12.24    & 42188/&13.46    & 42225/&13.38    & 45718/&6.21      \\
                             &  4 & 8-&(28555.40)& 53567/&$-$87.59  & 31423/&$-$10.04& 26983/&5.51     & 25993/&8.97     & 26630/&6.74     & 30107/&$-$5.43   \\
                             &  5 & 8-&(31701.46)& 56726/&$-$78.94  & 34429/&$-$8.60 & 30171/&4.83     & 29125/&8.13     & 29897/&5.69     & 33402/&$-$5.36   \\
                             &  4 & 7-&(28818.44)& 54136/&$-$87.85  & 32096/&$-$11.37& 27392/&4.95     & 26549/&7.87     & 27125/&5.88     & 30584/&$-$6.12   \\
                             &  5 & 7-&(29610.99)& 54836/&$-$85.19  & 32698/&$-$10.43& 28183/&4.82     & 27212/&8.10     & 27868/&5.89     & 31307/&$-$5.73   \\
                             &  6 & 7-&(32559.55)& 58105/&$-$78.46  & 36087/&$-$10.83& 31473/&3.34     & 30569/&6.11     & 31086/&4.52     & 34561/&$-$6.15   \\
                             &  7 & 7-&(36636.87)& 59905/&$-$63.51  & 37682/&$-$2.85 & 33316/&9.07     & 32320/&11.78    & 32975/&9.99     & 36465/&0.47      \\
                             &  1 & 3-&(29466.42)& 54751/&$-$85.81  & 32737/&$-$11.10& 27882/&5.38     & 27240/&7.56     & 27608/&6.31     & 31075/&$-$5.46   \\
                             &  2 & 3-&(31846.16)& 57174/&$-$79.53  & 35181/&$-$10.47& 30358/&4.67     & 29715/&6.69     & 30072/&5.57     & 33550/&$-$5.35   \\
                             &  3 & 3-&(33185.64)& 60105/&$-$81.12  & 38387/&$-$15.67& 33131/&0.17     & 32658/&1.59     & 32569/&1.86     & 36060/&$-$8.66   \\
                             &  4 & 3-&(36167.30)& 62975/&$-$74.12  & 41433/&$-$14.56& 35937/&0.64     & 35629/&1.49     & 35597/&1.58     & 39120/&$-$8.16   \\
                             &  5 & 3-&(37812.87)& 64373/&$-$70.24  & 42616/&$-$12.70& 37526/&0.76     & 36983/&2.20     & 36919/&2.36     & 40418/&$-$6.89   \\
                             &  6 & 3-&(38924.30)& 65174/&$-$67.44  & 43483/&$-$11.71& 38333/&1.52     & 37679/&3.20     & 37621/&3.35     & 41125/&$-$5.65   \\
                             &  7 & 3-&(40407.72)& 66462/&$-$64.48  & 44850/&$-$10.99& 39574/&2.06     & 39111/&3.21     & 38966/&3.57     & 42478/&$-$5.12   \\
                             &  1 & 2-&(38563.97)& 57758/&$-$49.77  & 35723/&7.37    & 30923/&19.81    & 30267/&21.51    & 30638/&20.55    & 34144/&11.46     \\
\hline
\enddata 
\tablecomments{The relative difference compared with NIST/(SE) data is given in percent.}
\end{deluxetable*}

\begin{figure}[ht!]
\includegraphics[width=0.47\textwidth]{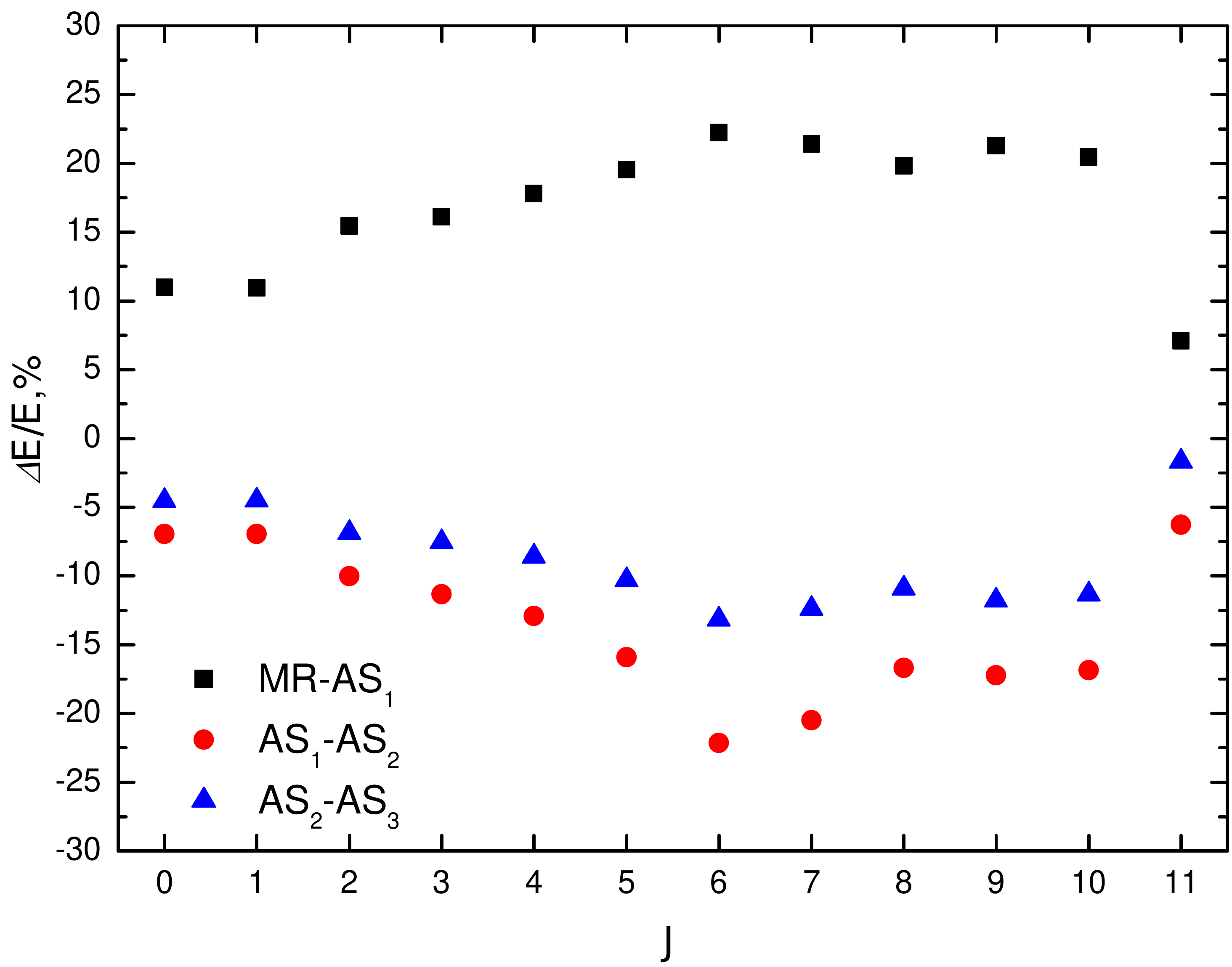}
\caption{Convergence of the lowest states of the $4f^{11}5d$ configuration with $J=0-11$ in the energy spectrum (\textbf{SD 4f} strategy).}
\label{lJkonvergensy}
\end{figure}

\begin{figure}[ht!]
\includegraphics[width=0.47\textwidth]{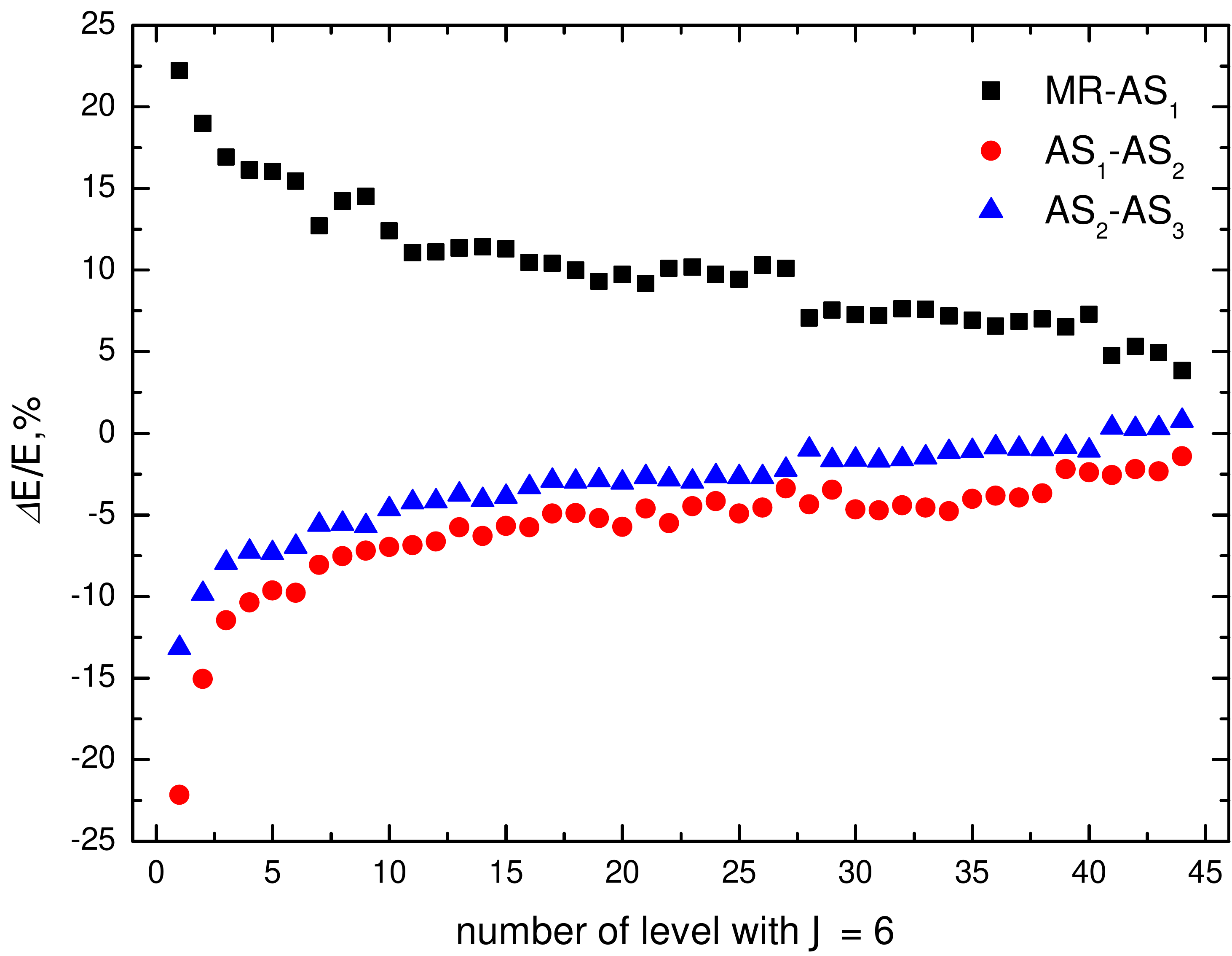}	
\caption{Convergence of all energy values of the $4f^{11}5d$ configuration with $J=6$ (\textbf{SD 4f} strategy).}
\label{J6konvergensy}
\end{figure}

The convergence of the obtained energies was evaluated by the following equation 
$\Delta E/E = (E_{AS_{N-1}}-E_{AS_{N}})/E_{AS_{N-1}}$. 
The relative difference ($\overline{\Delta E/E} =\frac{\sum|\Delta E_{i}/E_i|}{N}$) 
between active space $AS_2$ and $AS_3$ using the \textbf{SD 4f} strategy (when all 399 levels are included) is about 2.6\%. 
By analyzing the results we observe that energies for some $J$ values converge much faster than for others.
This is seen from Figure \ref{lJkonvergensy}, where the convergence for the lowest states of the 
$4f^{11}5d$ configuration with $J=0-11$ is presented. 
For example, the difference between $AS_2$ and $AS_3$ for $J=0$ is about 5\% while for lowest state with $J=6$ it reaches 13\%. 
After the studies of energy levels with different $J$ values, we observed that the lower energy levels converge much slower than the higher energy levels for a fixed $J$ value 
(see Figure \ref{J6konvergensy}).
From the Figure we see that even the third level converges much faster than the first one and the agreement between the energies for the last two active spaces is up to 0.3\%.
In conclusion, the active space has inconsiderable influence on the higher levels as compared to the lowest ones. The upper levels converge much faster.

\begin{figure}[ht!]
\includegraphics[width=0.47\textwidth]{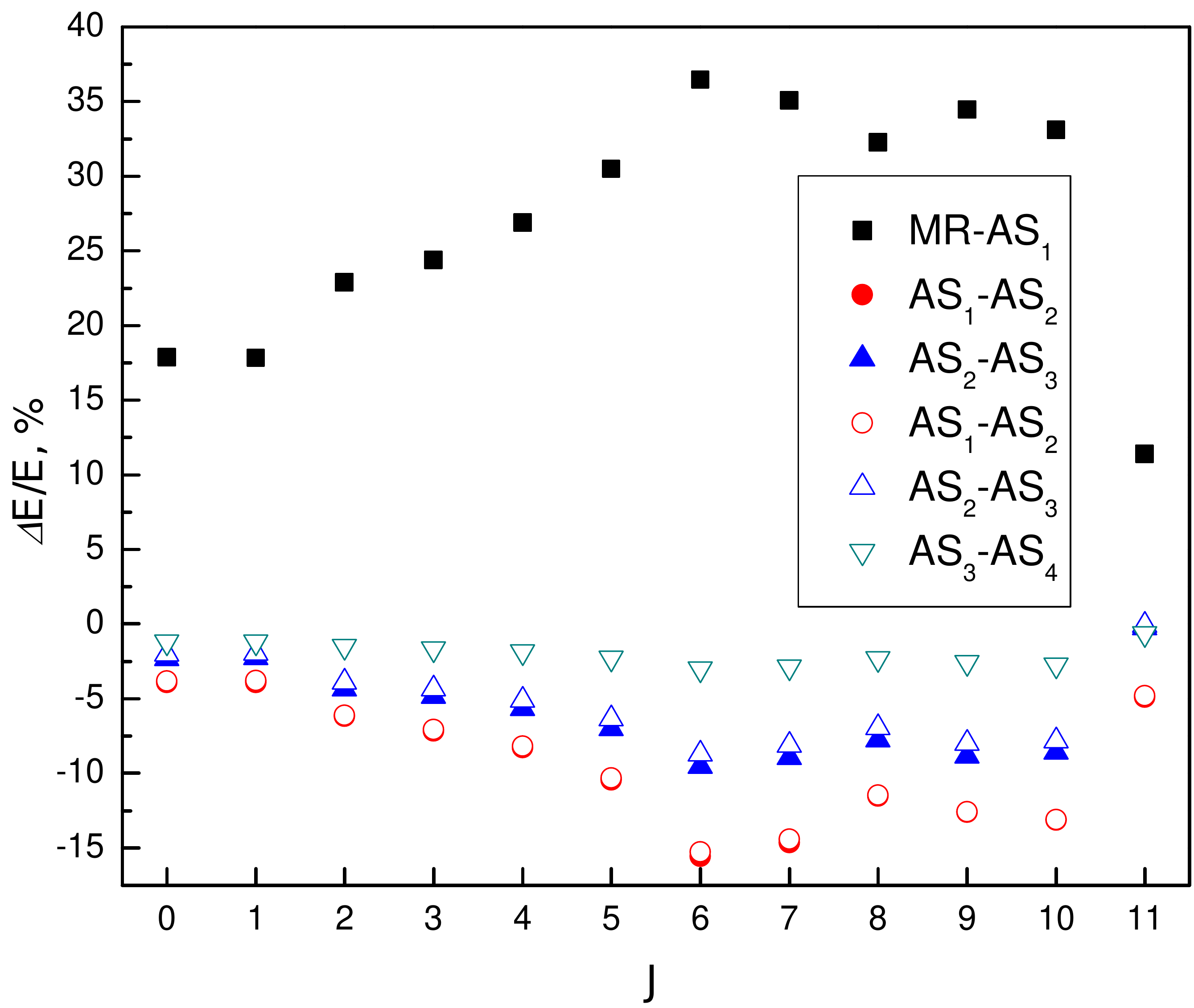}
\caption{Convergence of the lowest states of $4f^{11}5d$ configuration with $J=0,11$ in the energy spectrum. Results are obtained using the \textbf{SD 5d} strategy (open symbols mark the results when the \textbf{ZF$^{MCDHF}$} approach at $AS_{2,3,4}$ is applied).}
\label{lJkonvergensy_SD_5d}
\end{figure}

\begin{figure}[ht!]
\includegraphics[width=0.47\textwidth]{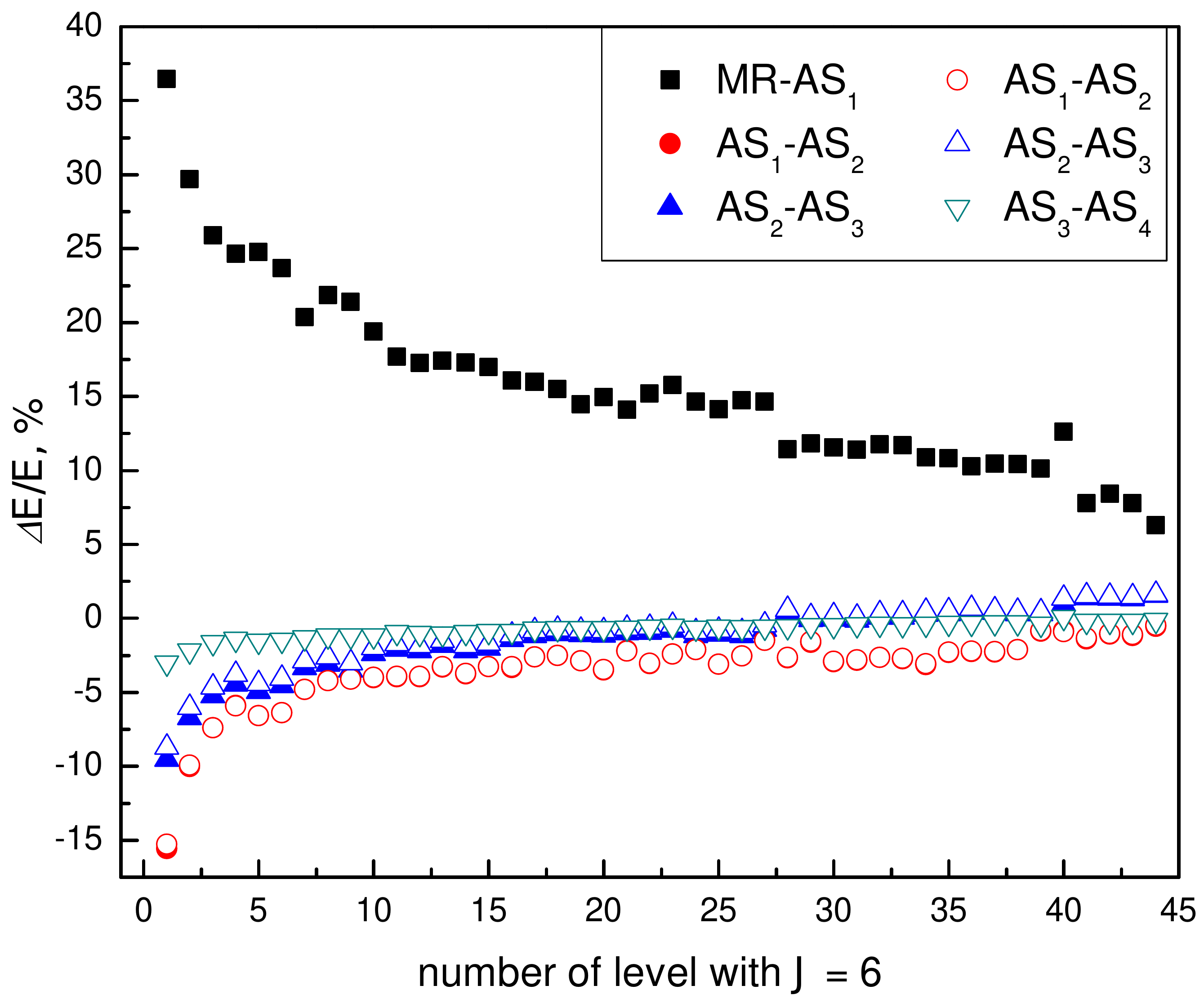}
\caption{Convergence of all energy values of the $4f^{11}5d$ configuration with $J=6$. Results are obtained using the \textbf{SD 5d} strategy (open symbols mark the results when the \textbf{ZF$^{MCDHF}$} approach at $AS_{2,3,4}$ is applied).}
\label{J6konvergensy_SD_5d}
\end{figure}

The lowest levels according to Hund's first rule have the largest multiplicity. 
For a given set of eigenstates, the lowest state will have largest multiplicity. 
Almost all the lowest levels for each $J$ in case of the $4f^{11}5d$ configuration have the largest multiplicity (except $J=9$), 
and all these levels converge slower than the higher ones (as it can be seen from Figures \ref{J6konvergensy} and \ref{J6konvergensy_SD_5d}). 
However, even in the set of levels with the largest multiplicity, a large differences in convergence is observable (see Figures \ref{lJkonvergensy} and \ref{lJkonvergensy_SD_5d}). 
From these Figures it can also be seen that the CSFs from the $AS_{1}$ (black squares) have the largest influence. 
The first active space has a larger influence on energy levels in the \textbf{SD 5d} strategy  than in the \textbf{SD 4f} strategy.

The results of the \textbf{SD 4f} strategy substantially disagree with NIST/(SE) data (see Table \ref{wf}) 
for states of the $4f^{11}5d$ configuration, and after adding one more layer ($AS_3$) to the computations, the disagreement increases. 
From the Table it is seen that after including substitutions from the $5d$ shell (\textbf{SD 5d} strategy) the results agree much better.
The averaged uncertainty of obtained results from the \textbf{SD 5d} strategy at $AS_2$ is around 5.6\% comparing with NIST or SE data.
By studying the convergence of the results obtained using the \textbf{SD 5d} strategy we see similar trends as those from the \textbf{SD 4f} strategy.
Firstly, energies for different $J$ values converge differently.
Secondly, lower energy levels converge much slower than the higher energy levels.
But in case of the \textbf{SD 5d} strategy the energies converge much faster comparing with the \textbf{SD 4f} strategy (see Figures \ref{lJkonvergensy_SD_5d} and \ref{J6konvergensy_SD_5d}).
For example, the difference between $AS_2$ and $AS_3$ for $J=0$ is about 2.3\% and 9.5\% for lowest state with $J=6$.

\subsection{Studies of core-valence and core-core electron correlations}
\label{CC_CV_results}
The investigations of core-core and core-valence electron correlations contributions to the transition energies are presented in Table \ref{wf2}.
From the Table it is seen that by including substitutions just from the valence shell ($4f$) and core shells ($5p$) or ($5s$)
(\textbf{SD 5p} or \textbf{SD 5s} strategy) the results are in worse agreement with NIST/(SE). In case of the \textbf{SD 5p} strategy this 
disagreement is very large.
The relative difference compared with NIST/(SE) data is reduced when substitutions from $4f$, $5d$ and $5p$ or $5s$ shells are allowed.
The averaged uncertainty of the obtained results from strategies 
\textbf{SrD 5p 5d}, 
\textbf{SD 5s 5d}, 
\textbf{SrD 5s 5p 5d}, 
\textbf{SD 5s 5p 5d} is similar, around 5-7\% comparing with NIST or SE data.
As was mentioned above, inclusion of the substitutions from the core shells ($5p$ or $5s$) increases
the number of CSFs dramatically (see Table \ref{summary}).
So for further investigations substitutions from the $5p$ and $5s$ shells were neglected. 

\subsection{Optimal strategy for electron correlations}
The \textbf{SD 5d} strategy was chosen as the optimal strategy 
considering achieved accuracy of the results and the computational resources needed for the calculations.
The main goal of this work is to obtain accurate energy levels of the ground and first excited configurations of Er$^{2+}$. 
So we give priority to balanced electron correlation effects which improves the energy separations.

\subsection{Impact of the zero-first-order method}
The ZF method 
was applied at different stages of the calculations to reduce computation resources, as it was described in Section \ref{ZF_computation}.
The impact of the ZF method was studied using the \textbf{SD 4f} and \textbf{SD 5d} strategies.
In the investigations of the effect of ZF on the energy levels all 399 states were included. 
The zero-first-order method (see \textbf{SD 4f ZF$^{MCDHF}$} column in Table \ref{ZF_contribution}) 
has up to 0.08\% impact on the values of the energy levels at $AS_3$ if all levels are compared.
From Table \ref{ZF_contribution_core} we see that the ZF method for MCDHF calculations (see \textbf{SD 5d ZF$^{MCDHF}$} column)
affects on average the values of the energy levels at $AS_3$ by 0.29\%, and in some cases up to~1.01\% .

The application of the ZF method for the RCI computation only (see \textbf{SD 4f ZF$_{RCI}$} column in Table \ref{ZF_contribution}), 
has a larger influence on the energy levels; it is up to 2.84\% at $AS_3$ and 1.61\% in average for all states. Using 
the \textbf{SD 5d ZF$_{RCI}$} strategy
(see Table \ref{ZF_contribution_core}) the contribution of ZF in RCI is up to 2.69\% at $AS_3$ and 1.45\% in average for all states.

When the ZF method was applied for the RCI computations using orbitals from \textbf{SD 4f ZF$^{MCDHF}$} 
the energies changed in average about 0.5\% (\textbf{SD 4f ZF$^{MCDHF}_{RCI}$}) and up to 2.39\% for some levels.
Using the \textbf{SD 5d ZF$^{MCDHF}_{RCI}$} strategy (see Table \ref{ZF_contribution_core}) 
the influence of the ZF method is up to 3.95\% at $AS_3$ and 0.66\% in average for all states.

From the above study we infer that the impact of the ZF order method on the energy levels is very small in self consistent field computations for both strategies.
In the case of the \textbf{SD 5d ZF$^{MCDHF}$} strategy, the effect on the energy levels at $AS_3$ is only 0.29\%.

\begin{deluxetable*}{lrlrrrrrrrrrrrrrrrrr}[ht!!]
\tabletypesize{\tiny}
\tablecaption{\label{ZF_contribution} Energy Levels from RCI Calculations Using the ZF Approach in Different Steps of the Calculations (\textbf{SD 4f} Strategy).}
\tablehead      {
$LS$& POS &\multicolumn{1}{c}{JP}&\multicolumn{1}{c}{NIST/(SE)} &\multicolumn{2}{c}{\textbf{SD 4f}} &&\multicolumn{2}{c}{\textbf{SD 4f ZF$^{MCDHF}$}}                                                                                                
&&\multicolumn{2}{c}{\textbf{SD 4f ZF$_{RCI}$}}&&\multicolumn{3}{c}{\textbf{SD 4f ZF$^{MCDHF}_{RCI}$}} \\
\cline{5-6} \cline{8-9}\cline{11-12} \cline{14-16}                 
&&&                                        &$AS_2$ &$AS_3$ &&$AS_2$ &$AS_3$ &&$AS_2$ &$AS_3$  &&$AS_2$ &$AS_3$ & $AS_4$
}
\startdata
$4f^{12}~{}^3H$               & 1 & 6+&     0.00 &     0 &     0 &&     0 &  0.00 &&     0 &     0  &&  0    & 0     & 0     \\
$4f^{12}~{}^3F$               & 1 & 4+&  5081.79 &  5894 &  5744 &&  5895 &  5744 &&  5834 &  5628  && 5832  & 5617  & 5572  \\
$4f^{12}~{}^3H$               & 1 & 5+&  6969.78 &  6784 &  6805 &&  6786 &  6806 &&  6790 &  6791  && 6794  & 6796  & 6798  \\
$4f^{12}~{}^3H$               & 2 & 4+& 10785.48 & 10957 & 10889 && 10961 & 10889 && 10912 & 10786  && 10915 & 10785 & 10766 \\
$4f^{12}~{}^3F$               & 1 & 3+&(12472.55)& 13565 & 13282 && 13570 & 13279 && 13472 & 13076  && 13475 & 13066 & 13019 \\
$4f^{12}~{}^3F$               & 1 & 2+&(13219.80)& 14824 & 14446 && 14829 & 14442 && 14777 & 14290  && 14780 & 14280 & 14216 \\
$4f^{12}~{}^1G$               & 3 & 4+&(18383.59)& 18359 & 18381 && 18361 & 18388 && 18188 & 18116  && 18187 & 18108 & 18053 \\
$4f^{11}~{}(^4I^1)~{}5d~{}^5G$& 1 & 6-& 16976.09 & 18983 & 21480 && 18976 & 21466 && 19666 & 20919  && 19634 & 22004 & 23585 \\
$4f^{11}~{}(^4I^1)~{}5d~{}^5H$& 1 & 7-& 17647.76 & 19877 & 22337 && 19875 & 22323 && 20543 & 21751  && 20513 & 22839 & 24421 \\
$4f^{11}~{}(^4I^1)~{}5d~{}^3L$& 1 & 9-& 18976.74 & 20916 & 23382 && 20933 & 23369 && 21593 & 22753  && 21577 & 23859 & 25438 \\
$4f^{11}~{}(^4I^1)~{}5d~{}^5I$& 1 & 8-& 19918.17 & 22273 & 24699 && 22285 & 24685 && 22923 & 24060  && 21577 & 25163 & 26746 \\
$4f^{11}~{}(^4I^1)~{}5d~{}^5L$& 1 &10-& 20470.13 & 21673 & 24130 && 21692 & 24116 && 22306 & 23463  && 22293 & 24575 & 26146 \\
$4f^{11}~{}(^4I^1)~{}5d~{}^5K$& 2 & 9-& 21688.17 & 23664 & 26085 && 23687 & 26071 && 24285 & 25387  && 24275 & 26504 & 28084 \\
$4f^{11}~{}(^4I^1)~{}5d~{}^5G$& 1 & 5-& 22016.77 & 24164 & 26648 && 24167 & 26634 && 24936 & 26172  && 24906 & 27264 & 28844 \\
$4f^{11}~{}(^4I^1)~{}5d~{}^5H$& 2 & 6-& 22606.07 & 24963 & 27418 && 24970 & 27404 && 25701 & 26896  && 25675 & 27993 & 29572 \\
$4f^{11}~{}(^4I^1)~{}5d~{}^5I$& 2 & 8-& 22951.42 & 25085 & 27514 && 25111 & 27501 && 25736 & 26835  && 25726 & 27954 & 29535 \\
$4f^{11}~{}(^4I^1)~{}5d~{}^5I$& 2 & 7-& 23302.78 & 26039 & 28459 && 26061 & 28445 && 26723 & 27834  && 26710 & 28948 & 30529 \\
$4f^{11}~{}(^4I^1)~{}5d~{}^5L$& 3 & 8-& 25482.12 & 27453 & 29916 && 27481 & 29903 && 28211 & 29364  && 28198 & 30479 & 32061 \\
$4f^{11}~{}(^4I^1)~{}5d~{}^5H$& 3 & 5-& 26192.66 & 29109 & 31551 && 29122 & 31537 && 29877 & 31044  && 29853 & 32145 & 33715 \\
$4f^{11}~{}(^4I^1)~{}5d~{}^5K$& 3 & 7-& 26579.91 & 29220 & 31644 && 29245 & 31630 && 29967 & 31094  && 29952 & 32206 & 33781 \\
                             &  1 & 4-&(26648.59)& 28786 & 31252 && 28792 & 31238 && 29595 & 30816  && 29565 & 31908 & 33482 \\
                             &  2 & 4-&(29469.40)& 32112 & 34506 && 32121 & 34492 && 32857 & 33980  && 32830 & 35074 & 36623 \\
                             &  3 & 4-&(30750.22)& 34590 & 36669 && 34584 & 36655 && 35182 & 35938  && 35153 & 37015 & 38514 \\
                             &  4 & 4-&(32196.96)& 35818 & 38077 && 35831 & 38064 && 36504 & 37462  && 36482 & 38560 & 40091 \\
                             &  5 & 4-&(33033.10)& 36800 & 38990 && 36816 & 38976 && 37462 & 38323  && 37445 & 39432 & 40969 \\
                             &  6 & 4-&(35903.96)& 40430 & 42276 && 40430 & 42262 && 41007 & 41478  && 40981 & 42564 & 44050 \\
                             &  7 & 4-&(37608.12)& 41896 & 44068 && 41919 & 44054 && 42533 & 43359  && 42519 & 44467 & 45981 \\
                             &  8 & 4-&(39667.36)& 43809 & 45645 && 43809 & 45631 && 44370 & 44833  && 44343 & 45916 & 47388 \\
                             &  9 & 4-&(40580.40)& 44448 & 46293 && 44452 & 46279 && 44964 & 45424  && 44942 & 46513 & 48012 \\
                             & 10 & 4-&(46937.23)& 44848 & 46678 && 44856 & 46664 && 45435 & 45881  && 45412 & 46975 & 48464 \\
                             &  3 & 9-&(27471.61)& 28932 & 31400 && 28964 & 31386 && 29672 & 30820  && 29663 & 31942 & 33516 \\
                             &  3 & 6-&(27472.46)& 30355 & 34985 && 30374 & 32747 && 31102 & 32230  && 31083 & 33335 & 34907 \\
                             &  4 & 6-&(28777.74)& 32618 & 35063 && 32642 & 34972 && 33310 & 34342  && 33300 & 35463 & 37023 \\
                             &  5 & 6-&(30283.09)& 32668 & 36114 && 32697 & 35049 && 33393 & 34505  && 33378 & 35614 & 37176 \\
                             &  6 & 6-&(31095.82)& 33767 & 36114 && 33789 & 36100 && 34471 & 35516  && 34455 & 36623 & 38172 \\
                             &  7 & 6-&(33191.53)& 36889 & 38955 && 36909 & 38941 && 37476 & 38145  && 37462 & 39243 & 40744 \\
                             &  8 & 6-&(33875.19)& 38090 & 40207 && 38115 & 40194 && 38707 & 39459  && 38698 & 40581 & 42120 \\
                             &  9 & 6-&(35856.62)& 39697 & 41952 && 39734 & 41939 && 40377 & 41260  && 40371 & 42385 & 43931 \\
                             & 10 & 6-&(36570.10)& 40776 & 42673 && 40787 & 42659 && 41303 & 41822  && 41287 & 42926 & 44426 \\
                             &  3 & 5-&(27870.83)& 31235 & 33646 && 31267 & 33632 && 31956 & 33014  && 31948 & 34138 & 35701 \\
                             &  4 & 5-&(29995.62)& 33042 & 35388 && 33056 & 35374 && 33750 & 34811  && 33728 & 35909 & 37455 \\
                             &  5 & 5-&(31214.52)& 34758 & 36967 && 34768 & 36954 && 35395 & 36273  && 35374 & 37363 & 38877 \\
                             &  6 & 5-&(32614.37)& 35994 & 38025 && 35995 & 38011 && 36569 & 37289  && 36545 & 38380 & 39889 \\
                             &  7 & 5-&(33704.29)& 37449 & 39616 && 37468 & 39602 && 38095 & 38931  && 38079 & 40039 & 41571 \\
                             &  8 & 5-&(36330.81)& 40031 & 42203 && 40054 & 42189 && 40688 & 41505  && 40673 & 42564 & 44138 \\
                             &  9 & 5-&(36655.60)& 41618 & 43638 && 41638 & 43624 && 42201 & 42850  && 42189 & 43961 & 45469 \\
                             & 10 & 5-&(39265.81)& 44536 & 46335 && 44546 & 46321 && 45111 & 45507  && 45091 & 46601 & 48083 \\
                             & 11 & 5-&(40857.10)& 45058 & 46878 && 45063 & 46864 && 45575 & 46024  && 45554 & 47116 & 48588 \\
                             & 12 & 5-&(46552.18)& 45507 & 47406 && 45521 & 47393 && 46086 & 46602  && 46068 & 47703 & 49195 \\
                             & 13 & 5-&(48747.15)& 47313 & 49233 && 47330 & 49219 && 47881 & 48424  && 47865 & 49525 & 51002 \\
                             &  4 & 8-&(28555.40)& 30680 & 33109 && 30715 & 33095 && 31424 & 32525  && 31416 & 33650 & 35232 \\
                             &  5 & 8-&(31701.46)& 33681 & 36102 && 33717 & 36089 && 34407 & 35485  && 34399 & 36609 & 38170 \\
                             &  4 & 7-&(28818.44)& 31327 & 33744 && 31356 & 33731 && 32067 & 33172  && 32055 & 34289 & 35862 \\
                             &  5 & 7-&(29610.99)& 31953 & 34373 && 31987 & 34359 && 32699 & 33791  && 32689 & 34913 & 36491 \\
                             &  6 & 7-&(32559.55)& 35372 & 37743 && 35412 & 37729 && 36082 & 37093  && 36078 & 38221 & 39782 \\
                             &  7 & 7-&(36636.87)& 36973 & 39291 && 37007 & 39277 && 37649 & 38597  && 37642 & 39716 & 41249 \\
                             &  1 & 3-&(29466.42)& 31943 & 34353 && 31949 & 34339 && 32712 & 33871  && 32682 & 34961 & 36516 \\
                             &  2 & 3-&(31846.16)& 34404 & 36802 && 34415 & 36788 && 35149 & 36282  && 35123 & 37379 & 38932 \\
                             &  3 & 3-&(33185.64)& 37798 & 39691 && 37788 & 39677 && 38350 & 38906  && 38320 & 39982 & 41475 \\
                             &  4 & 3-&(36167.30)& 40736 & 42934 && 40763 & 42920 && 41406 & 42218  && 41394 & 43330 & 44849 \\
                             &  5 & 3-&(37812.87)& 42093 & 43938 && 42085 & 43923 && 42655 & 43141  && 42622 & 44215 & 45696 \\
                             &  6 & 3-&(38924.30)& 42968 & 44779 && 42969 & 44765 && 43507 & 43943  && 43484 & 45031 & 46529 \\
                             &  7 & 3-&(40407.72)& 44318 & 46177 && 44324 & 46163 && 44847 & 45302  && 44827 & 46395 & 47874 \\
                             &  1 & 2-&(38563.97)& 34941 & 37339 && 34952 & 37325 && 35680 & 36810  && 35654 & 37907 & 39458 \\		
\hline
\enddata 
\end{deluxetable*}

\begin{deluxetable*}{lrlrrrrrrrrrrrrrrrrr}[ht!!]
\tabletypesize{\tiny}
\tablecaption{\label{ZF_contribution_core} Energy Levels from RCI Calculations Using the ZF Approach in Different Steps of the Calculations (\textbf{SD 5d} Strategy).}
\tablehead      {
$LS$& POS &\multicolumn{1}{c}{JP}&\multicolumn{1}{c}{NIST/(SE)}
&\multicolumn{2}{c}{\textbf{SD 5d}}
&&\multicolumn{2}{c}{\textbf{SD 5d ZF$^{MCDHF}$}}
&&\multicolumn{2}{c}{\textbf{SD 5d ZF$_{RCI}$}}
&&\multicolumn{2}{c}{\textbf{SD 5d ZF$^{MCDHF}_{RCI}$}} \\                                          
\cline{5-6} \cline{8-9} \cline{11-12} \cline{14-15}
                           &&   &          & $AS_2$& $AS_3$&& $AS_2$& $AS_3$&&  $AS_2$&$AS_3$ &&  $AS_2$& $AS_3$ 
}
\startdata
$4f^{12}~{}^3H$               & 1 & 6+&     0.00 & 0     &     0 && 0	    & 0	    &&      0 &     0 && 0	    & 0	    \\
$4f^{12}~{}^3F$               & 1 & 4+&  5081.79 & 5895  & 5744  && 5895	& 5744	&&   5833 &  5628 && 5832	& 5617	\\
$4f^{12}~{}^3H$               & 1 & 5+&  6969.78 & 6784  &  6805 && 6786	& 6806	&&   6790 &  6791 && 6794	& 6796	\\
$4f^{12}~{}^3H$               & 2 & 4+& 10785.48 & 10958 & 10889 && 10961	& 10890	&&  10912 & 10786 && 10916	& 10786	\\
$4f^{12}~{}^3F$               & 1 & 3+&(12472.55)& 13566 & 13282 && 13570	& 13279	&&  13471 & 13075 && 13475	& 13066	\\
$4f^{12}~{}^3F$               & 1 & 2+&(13219.80)& 14825 & 14446 && 14829	& 14442	&&  14776 & 14290 && 14781	& 14280	\\
$4f^{12}~{}^1G$               & 3 & 4+&(18383.59)& 18360 & 18381 && 18361	& 18388	&&  18187 & 15728 && 18188	& 18109	\\
$4f^{11}~{}(^4I^1)~{}5d~{}^5G$& 1 & 6-& 16976.09 & 14673 & 16073 && 14638	& 15912	&&  15109 & 15728 && 15045	& 16734	\\
$4f^{11}~{}(^4I^1)~{}5d~{}^5H$& 1 & 7-& 17647.76 & 15613 & 17006 && 15583	& 16849	&&  16057 & 16642 && 15997	& 17652	\\
$4f^{11}~{}(^4I^1)~{}5d~{}^3L$& 1 & 9-& 18976.74 & 16725 & 18204 && 16721	& 18057	&&  17194 & 17734 && 17150	& 18766	\\
$4f^{11}~{}(^4I^1)~{}5d~{}^5I$& 1 & 8-& 19918.17 & 17988 & 19380 && 17977	& 19226	&&  18417 & 18924 && 18369	& 19951	\\
$4f^{11}~{}(^4I^1)~{}5d~{}^5L$& 1 &10-& 20470.13 & 17644 & 19158 && 17640	& 19019	&&  18088 & 18656 && 18048	& 19695	\\
$4f^{11}~{}(^4I^1)~{}5d~{}^5K$& 2 & 9-& 21688.17 & 19561 & 21006 && 19562	& 20861	&&  19985 & 20485 && 19947	& 21528	\\
$4f^{11}~{}(^4I^1)~{}5d~{}^5G$& 1 & 5-& 22016.77 & 19890 & 21285 && 19866	& 21125	&&  20456 & 21058 && 20396	& 22074	\\
$4f^{11}~{}(^4I^1)~{}5d~{}^5H$& 2 & 6-& 22606.07 & 20716 & 22102 && 20696	& 21944	&&  21256 & 21822 && 21199	& 22842	\\
$4f^{11}~{}(^4I^1)~{}5d~{}^5I$& 2 & 8-& 22951.42 & 20903 & 22321 && 20906	& 22170	&&  21357 & 21837 && 21318	& 22881	\\
$4f^{11}~{}(^4I^1)~{}5d~{}^5I$& 2 & 7-& 23302.78 & 21733 & 23103 && 21733	& 22946	&&  22215 & 22689 && 22172	& 23727	\\
$4f^{11}~{}(^4I^1)~{}5d~{}^5L$& 3 & 8-& 25482.12 & 23276 & 24757 && 23283	& 24610	&&  23839 & 24376 && 23797	& 25416	\\
$4f^{11}~{}(^4I^1)~{}5d~{}^5H$& 3 & 5-& 26192.66 & 24808 & 26162 && 24796	& 26001	&&  25368 & 25881 && 25315	& 26907	\\
$4f^{11}~{}(^4I^1)~{}5d~{}^5K$& 3 & 7-& 26579.91 & 24959 & 26367 && 24963	& 26215	&&  25510 & 26005 && 25467	& 27043	\\
                             &  1 & 4-&(26648.59)& 24556 & 25947 && 24533	& 25789	&&  25166 & 25762 && 25104	& 26776	\\
                             &  2 & 4-&(29469.40)& 27877 & 29198 && 27858	& 29040	&&  28423 & 28910 && 28365	& 29926	\\
                             &  3 & 4-&(30750.22)& 30274 & 31299 && 30242	& 31141	&&  30648 & 30782 && 30588	& 31783	\\
                             &  4 & 4-&(32196.96)& 31408 & 32609 && 31401	& 32446	&&  31894 & 32225 && 31845	& 33250	\\
                             &  5 & 4-&(33033.10)& 32374 & 33454 && 32361	& 33291	&&  32814 & 33018 && 32764	& 34044	\\
                             &  6 & 4-&(35903.96)& 36133 & 36946 && 36106	& 36791	&&  36501 & 36350 && 36444	& 37359	\\
                             &  7 & 4-&(37608.12)& 37438 & 38533 && 37440	& 38367	&&  37872 & 38055 && 37829	& 39088	\\
                             &  8 & 4-&(39667.36)& 39556 & 40364 && 39530	& 40209	&&  39911 & 39765 && 39854	& 40770	\\
                             &  9 & 4-&(40580.40)& 40165 & 41000 && 40145	& 40846	&&  40463 & 40318 && 40412	& 41330	\\
                             & 10 & 4-&(46937.23)& 40498 & 41266 && 40482	& 41108	&&  40866 & 40650 && 40815	& 41666	\\
                             &  3 & 9-&(27471.61)& 24896 & 26416 && 24906	& 26274	&&  25453 & 26006 && 25416	& 27055	\\
                             &  3 & 6-&(27472.46)& 26082 & 27443 && 26077	& 27287	&&  26623 & 27115 && 26575	& 28144	\\
                             &  4 & 6-&(28777.74)& 28112 & 29360 && 28123	& 29190	&&  28598 & 28957 && 28561	& 30003	\\
                             &  5 & 6-&(30283.09)& 28454 & 29857 && 28453	& 29708	&&  29009 & 29507 && 28963	& 30539	\\
                             &  6 & 6-&(31095.82)& 29549 & 30894 && 29548	& 30745	&&  30058 & 30480 && 30013	& 31510	\\
                             &  7 & 6-&(33191.53)& 32635 & 33718 && 32634	& 33567	&&  33025 & 33092 && 32982	& 34114	\\
                             &  8 & 6-&(33875.19)& 33618 & 34659 && 33620	& 34495	&&  34014 & 34125 && 33976	& 35170	\\
                             &  9 & 6-&(35856.62)& 35430 & 36666 && 35446	& 36512	&&  35906 & 36165 && 35871	& 37218	\\
                             & 10 & 6-&(36570.10)& 36480 & 37345 && 36467	& 37193	&&  36781 & 36644 && 36736	& 37669	\\
                             &  3 & 5-&(27870.83)& 26706 & 27992 && 26714	& 27823	&&  27225 & 27635 && 27185	& 28678	\\
                             &  4 & 5-&(29995.62)& 28760 & 30057 && 28749	& 29902	&&  29261 & 29690 && 29210	& 30711	\\
                             &  5 & 5-&(31214.52)& 30474 & 31638 && 30459	& 31484	&&  30890 & 31115 && 30839	& 32127	\\
                             &  6 & 5-&(32614.37)& 31714 & 32739 && 31692	& 32585	&&  32095 & 32222 && 32041	& 33237	\\
                             &  7 & 5-&(33704.29)& 33073 & 34182 && 33069	& 34022	&&  33511 & 33715 && 33466	& 34747	\\
                             &  8 & 5-&(36330.81)& 35735 & 36870 && 35735	& 36714	&&  36193 & 36386 && 36149	& 37420	\\
                             &  9 & 5-&(36655.60)& 37138 & 38054 && 37135	& 37888	&&  37503 & 37468 && 37461	& 38502	\\
                             & 10 & 5-&(39265.81)& 40220 & 41002 && 40208	& 40847	&&  40562 & 40325 && 40514	& 41344	\\
                             & 11 & 5-&(40857.10)& 40672 & 41476 && 40651	& 41317	&&  40976 & 40841 && 40924	& 41852	\\
                             & 12 & 5-&(46552.18)& 41210 & 42034 && 41200	& 41878	&&  41583 & 41409 && 41536	& 42432	\\
                             & 13 & 5-&(48747.15)& 42975 & 43871 && 42968	& 43715	&&  43336 & 43270 && 43291	& 44295	\\
                             &  4 & 8-&(28555.40)& 26563 & 28010 && 26576	& 27863	&&  27113 & 27607 && 27076	& 28658	\\
                             &  5 & 8-&(31701.46)& 29611 & 31074 && 29626	& 30931	&&  30149 & 30624 && 30114	& 31675	\\
                             &  4 & 7-&(28818.44)& 27034 & 28431 && 27041	& 28278	&&  27583 & 28061 && 27542	& 29102	\\
                             &  5 & 7-&(29610.99)& 27752 & 29159 && 27763	& 29008	&&  28303 & 28778 && 28264	& 29825	\\
                             &  6 & 7-&(32559.55)& 31168 & 32535 && 31186	& 32385	&&  31672 & 32063 && 31639	& 33116	\\
                             &  7 & 7-&(36636.87)& 32864 & 34212 && 32877	& 34067	&&  33355 & 33689 && 33320	& 34734	\\
                             &  1 & 3-&(29466.42)& 27718 & 29057 && 27694	& 28899	&&  28290 & 28822 && 28228	& 29834	\\
                             &  2 & 3-&(31846.16)& 30216 & 31554 && 30200	& 31397	&&  30772 & 31282 && 30715	& 32303	\\
                             &  3 & 3-&(33185.64)& 33433 & 34279 && 33397	& 34120	&&  33758 & 33701 && 33697	& 34700	\\
                             &  4 & 3-&(36167.30)& 36191 & 37278 && 36197	& 37110	&&  36645 & 36799 && 36603	& 37835	\\
                             &  5 & 3-&(37812.87)& 37805 & 38620 && 37771	& 38464	&&  38168 & 38032 && 38105	& 39027	\\
                             &  6 & 3-&(38924.30)& 38621 & 39395 && 38597	& 39238	&&  38938 & 38743 && 38883	& 39753	\\
                             &  7 & 3-&(40407.72)& 39948 & 40754 && 39930  & 40596 &&  40253 & 40068 && 40203  & 41083 \\ 
                             &  1 & 2-&(38563.97)& 30760 & 32092 && 30742	& 31935	&&  31306 & 31815 && 31248	& 32835	\\	
\hline
\enddata 
\end{deluxetable*}

\subsection{Final results}
\label{section_final_results}
Based on the analysis made in previous sections, the \textbf{SD 5d}
strategy was chosen as the optimal strategy.
Therefore this strategy with the orbitals taken from the \textbf{ZF$^{MCDHF}$} strategy  
was used to continue computations in $AS_4$ basis.
The final results of the present work are displayed in Table \ref{energy_final} together with NIST and SE data.
In first column of the Table we give identifications of energy levels in $LS$ or $JJ$ (see definition in \citet{Gaigalas_coupling}, Eq. (10) and (16)) coupling from our computations, 
in second column identifications of energy levels are from \citet{Wyart}. Labels in $LS$ coupling agree with identification given in the NIST database. Labels in $JJ$ coupling are given only for the part of the energy spectra 
that is used for the comparison with the results of Wyart et al.
The averaged uncertainty of the computed energy levels is 5.24\%, 2.68\%, 
respectively for states of the ground and excited configurations (see Table \ref{energy_final} \textbf{SD 5d ZF$^{MCDHF}$} strategy $AS_4$).
Root-mean-square (rms) deviations of these results for states of the ground and excited configurations from the NIST/(SE) data
are 649 cm$^{-1}$, and 1571 cm$^{-1}$, respectively.
If the ZF method is used in both the MCDHF and RCI calculations (\textbf{SD 5d ZF$^{MCDHF}_{RCI}$} strategy) 
the obtained data are in worse agreement (moderately about 7\%) with NIST or SE data. 

\begin{figure}[ht!]
\includegraphics[width=0.47\textwidth]{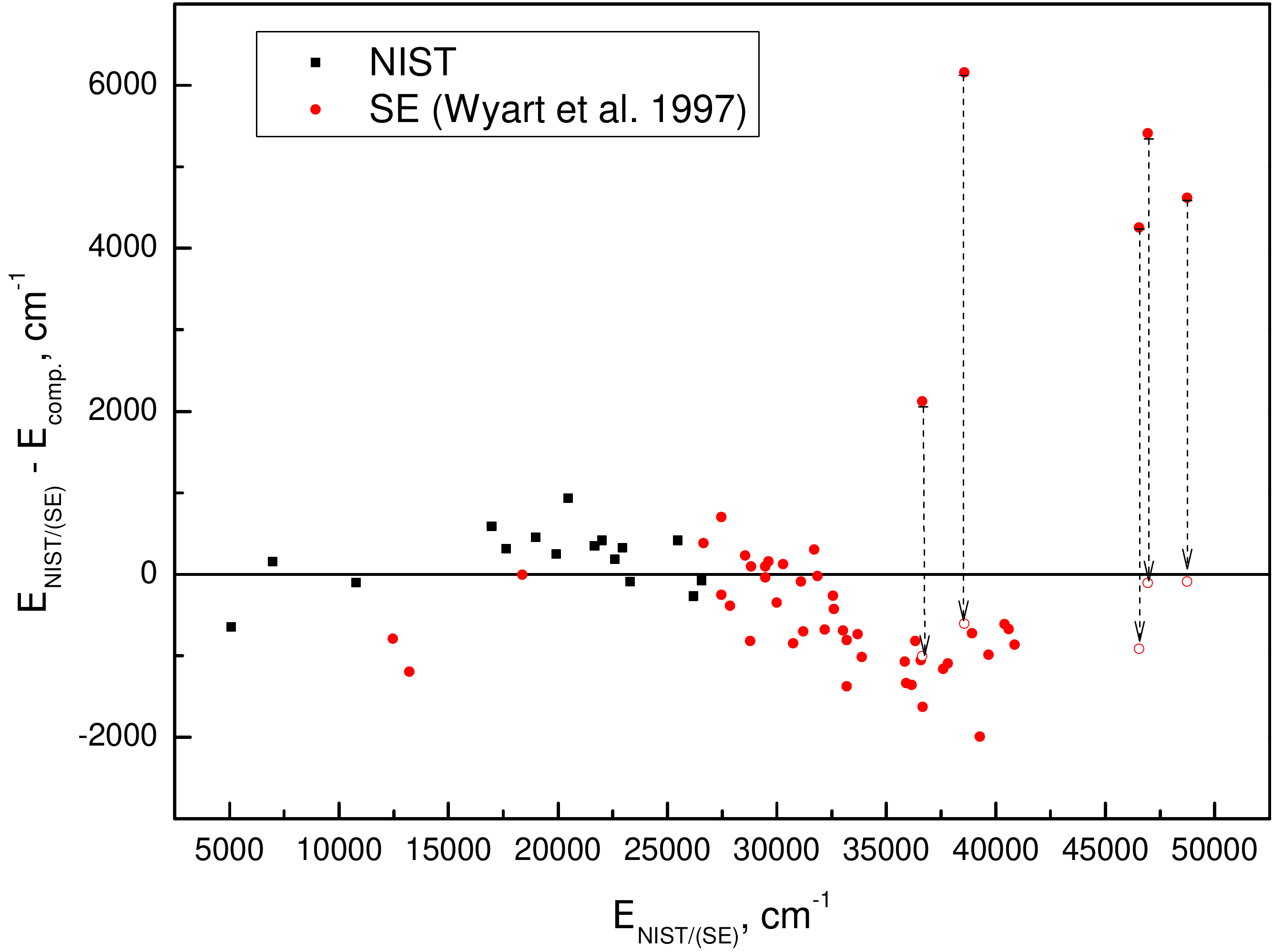}
\caption{A comparison of energy levels between the NIST or SE values \citet{Wyart} and results of the present study. The dashed arrows indicate the improved agreement resulting from a re-identification of the levels in \citet{Wyart}, see text for details.
} 
\label{energy_comparison_geras}
\end{figure}

Figure \ref{energy_comparison_geras} displays the differences 
between the NIST/(SE) energies and final results of the present study.
As it can be seen from Figure \ref{energy_comparison_geras}
and Table \ref{energy_final} (energy levels marked in gray color), there is a significant disagreement between states with the following identifications $J$=4 Pos=10, $J$=5 Pos 12, $J$=5 Pos=13, $J$=7 Pos=7, and $J$=2 Pos=1. 
Energy differences exceed 2000 cm$^{-1}$ for these five energy levels.
It is highly probable that the obtained differences result from
incorrect ordering and incomplete 
identification of energy levels presented by \citet{Wyart}.
Only for one level ($J$=7 Pos=7) from the five above mentioned levels \citet{Wyart} give identification 
in $JJ$ coupling, for the four others only configurations are given. The level is identified as $^4F_{9/2}~5d_{5/2}$ ($J$=7). 
We have transformed ASFs from $LS$ to $JJ$ coupling using the \textsc{Coupling} program developed by \citet{Gaigalas_coupling}.
The level $J$=7 Pos=7 has the $4f^{11}(^4I_{9/2})~5d_{5/2}~(9/2,5/2)$ label in $JJ$ coupling which disagree with Wyart et al.
By looking at levels which match the identification given by Wyart et al. we see that there is a fit for $J$=7 Pos=8 with identification $4f^{11}(^4F_{9/2})~5d_{5/2}~(9/2,5/2)$.
If we replace computed energy levels marked in gray color in Table \ref{energy_final}  
by energy levels suggested in Table \ref{energy_final_2} (presented by open red circles in Figure \ref{energy_final}), agreement with the 
NIST/(SE) data is much better.
The change in the differences between the NIST/(SE) energies and our final results
is shown by dashed arrows in Figure \ref{energy_comparison_geras}.
The rms deviation for states of the excited configuration 
(when five of the computed energy levels are replaced) is now only 747 cm$^{-1}$.
By comparing the labels of the levels for which Wyart et al. gives the full identification with our identification in $JJ$ coupling, the labels from both studies agree except for the levels (namely
$J$=4 Pos=8, $J$=6 Pos 5, $J$=6 Pos=10, $J$=5 Pos=3, and $J$=3 Pos=1). Level $J$=4 Pos=8 in the present work has the 
$4f^{11}(^4F_{7/2})~5d_{5/2}~(7/2,5/2)$ idendification; $J$=6 Pos 5 -- $4f^{11}(^4I_{9/2})~5d_{3/2}~(9/2,3/2)$; 
$J$=6 Pos=10 -- $4f^{11}(^4F_{9/2})~5d_{5/2}~(9/2,5/2)$; $J$=5 Pos=3 -- $4f^{11}(^4I_{15/2})~5d_{5/2}~(15/2,5/2)$; 
and $J$=3 Pos=1 -- $4f^{11}(^4I_{11/2})~5d_{5/2}~(11/2,5/2)$.
It was observed that the identification given in \citep{Wyart} for level $J$=3 Pos=1 is incorrect. That level was assigned as $^4I_{11/2}~5d_{3/2}$ but such a label for $J$=3 is not consistent with the selection rules.
The deeper analysis of uncertainties estimation is complicated 
because complete identification of energy levels was not given in the paper by \citet{Wyart}.

\begin{deluxetable*}{llrlrr@{\hskip 0.05pt}rrr@{\hskip 0.05pt}r}[ht!!]
\tabletypesize{\tiny}
\tablecaption{\label{energy_final} Comparison of Energy Levels from the Present Calculations Based on the \textbf{SD 5d} Strategy and Using the ZF Approach with NIST/SE data.}
\tablehead      {
$LS$/$JJ$& label in \cite{Wyart} & POS &\multicolumn{1}{c}{JP}&\multicolumn{1}{c}{NIST/(SE)}
&\multicolumn{2}{c}{\textbf{SD 5d ZF$^{MCDHF}$}}
&&\multicolumn{2}{c}{\textbf{SD 5d ZF$^{MCDHF}_{RCI}$}} \\                                          
\cline{6-7} \cline{9-10}
                           &&&   &          &\multicolumn{2}{c}{$AS_4$}  && \multicolumn{2}{c}{$AS_4$}
}
\startdata
$4f^{12}~{}^3H$                &                                           & 1 & 6+&     0.00 & 0    &         && 0        \\        
$4f^{12}~{}^3F$                &                                           & 1 & 4+&  5081.79 &  5731/&$-$12.77&& 5572 /&$-$9.65   \\
$4f^{12}~{}^3H$                &                                           & 1 & 5+&  6969.78 &  6814/&2.24    && 6798 /&2.47      \\
$4f^{12}~{}^3H$                &                                           & 2 & 4+& 10785.48 & 10890/&$-$0.97 && 10766/&0.18      \\
$4f^{12}~{}^3F$                &                                           & 1 & 3+&(12472.55)& 13267/&$-$6.37 && 13019/&$-$4.38   \\
$4f^{12}~{}^3F$                &                                           & 1 & 2+&(13219.80)& 14418/&$-$9.06 && 14216/&$-$7.54   \\
$4f^{12}~{}^1G$                &                                           & 3 & 4+&(18383.59)& 18393/&$-$0.05 && 18053/&1.80      \\
$4f^{11}~{}(^4I^1)~{}5d~{}^5G$ &                                           & 1 & 6-& 16976.09 & 16391/&3.44    && 18186/&$-$7.13   \\
$4f^{11}~{}(^4I^1)~{}5d~{}^5H$ &                                           & 1 & 7-& 17647.76 & 17333/&1.78    && 19110/&$-$8.29   \\
$4f^{11}~{}(^4I^1)~{}5d~{}^3L$ &                                           & 1 & 9-& 18976.74 & 18524/&2.39    && 20211/&$-$6.51   \\
$4f^{11}~{}(^4I^1)~{}5d~{}^5I$ &                                           & 1 & 8-& 19918.17 & 19672/&1.24    && 21401/&$-$7.44   \\
$4f^{11}~{}(^4I^1)~{}5d~{}^5L$ &                                           & 1 &10-& 20470.13 & 19537/&4.56    && 21155/&$-$3.34   \\
$4f^{11}~{}(^4I^1)~{}5d~{}^5K$ &                                           & 2 & 9-& 21688.17 & 21342/&1.60    && 22986/&$-$5.98   \\
$4f^{11}~{}(^4I^1)~{}5d~{}^5G$ &                                           & 1 & 5-& 22016.77 & 21604/&1.87    && 23527/&$-$6.86   \\
$4f^{11}~{}(^4I^1)~{}5d~{}^5H$ &                                           & 2 & 6-& 22606.07 & 22421/&0.82    && 24297/&$-$7.48   \\
$4f^{11}~{}(^4I^1)~{}5d~{}^5I$ &                                           & 2 & 8-& 22951.42 & 22629/&1.41    && 24329/&$-$6.00   \\
$4f^{11}~{}(^4I^1)~{}5d~{}^5I$ &                                           & 2 & 7-& 23302.78 & 23392/&$-$0.38 && 25176/&$-$8.04   \\
$4f^{11}~{}(^4I^1)~{}5d~{}^5L$ &                                           & 3 & 8-& 25482.12 & 25069/&1.62    && 26865/&$-$5.43   \\
$4f^{11}~{}(^4I^1)~{}5d~{}^5H$ &                                           & 3 & 5-& 26192.66 & 26464/&$-$1.03 && 28349/&$-$8.23   \\
$4f^{11}~{}(^4I^1)~{}5d~{}^5K$ &                                           & 3 & 7-& 26579.91 & 26661/&$-$0.31 && 28490/&$-$7.19   \\
$4f^{11}~(^4I_{11/2})~5d_{3/2}~(11/2,3/2)$  & $^4I_{11/2}~5d_{3/2}$         &  1 & 4-&(26648.59)& 26268/&1.43    && 28225/&$-$5.92   \\
$4f^{11}~(^4I_{11/2})~5d_{5/2}~(11/2,5/2)$  & $4f^{11}~5d$                  &  2 & 4-&(29469.40)& 29509/&$-$0.13 && 31348/&$-$6.37   \\
$4f^{11}~(^4F_{9/2})~5d_{3/2}~(9/2,3/2)$     & $^4F_{9/2}~5d_{3/2}$         &  3 & 4-&(30750.22)& 31599/&$-$2.76 && 33161/&$-$7.84   \\
$4f^{11}~(^4I_{9/2})~5d_{3/2}~(9/2,3/2)$     & $4f^{11}~5d$                 &  4 & 4-&(32196.96)& 32877/&$-$2.11 && 34655/&$-$7.63   \\
$4f^{11}~(^4F_{9/2})~5d_{3/2}~(9/2,3/2)$     & $4f^{11}~5d$                 &  5 & 4-&(33033.10)& 33728/&$-$2.10 && 35450/&$-$7.32   \\
$4f^{11}~(^4F_{7/2})~5d_{3/2}~(7/2,3/2)$     & $^4F_{9/2}~5d$               &  6 & 4-&(35903.96)& 37243/&$-$3.73 && 38722/&$-$7.85   \\
$4f^{11}~(^4I_{9/2})~5d_{5/2}~(9/2,5/2)$     & $^4I_{9/2}~5d_{5/2}$         &  7 & 4-&(37608.12)& 38774/&$-$3.10 && 40476/&$-$7.62   \\
$4f^{11}~(^4F_{7/2})~5d_{5/2}~(7/2,5/2)$     & $^4F_{7/2}~5d_{3/2}$         &  8 & 4-&(39667.36)& 40658/&$-$2.50 && 42126/&$-$6.20   \\
$4f^{11}~(^4S_{3/2})~5d_{5/2}~(3/2,5/2)$     & $4f^{11}~5d$                 &  9 & 4-&(40580.40)& 41258/&$-$1.67 && 42710/&$-$5.25   \\
$4f^{11}~(^4F_{7/2})~5d_{3/2}~(7/2,3/2)$     & $4f^{11}~5d$                 & \colorbox[rgb]{0.75,0.75,0.75}{10} & \colorbox[rgb]{0.75,0.75,0.75}{4-}&(46937.23)& \colorbox[rgb]{0.75,0.75,0.75}{41524/}&\colorbox[rgb]{0.75,0.75,0.75}{11.53}   && 43024/&8.34      \\
$4f^{11}~(^4I_{13/2})~5d_{5/2}~(13/2,5/2)$  & $^4I_{13/2}~5d_{5/2}$         &  3 & 9-&(27471.61)& 26769/&2.56    && 28508/&$-$3.77   \\
$4f^{11}~(^4I_{13/2})~5d_{3/2}~(13/2,3/2)$  & $^4I_{13/2}~5d_{3/2}$         &  3 & 6-&(27472.46)& 27729/&$-$0.93 && 29584/&$-$7.69   \\
$4f^{11}~(^4I_{13/2})~5d_{5/2}~(13/2,5/2)$  & $^4I_{13/2}~5d_{5/2}$         &  4 & 6-&(28777.74)& 29598/&$-$2.85 && 31434/&$-$9.23   \\
$4f^{11}~(^4I_{9/2})~5d_{3/2}~(9/2,3/2)$    & $^4I_{11/2}~5d_{3/2}$         &  5 & 6-&(30283.09)& 30162/&0.40    && 31975/&$-$5.59   \\
$4f^{11}~(^4I_{9/2})~5d_{3/2}~(9/2,3/2)$    & $^4I_{9/2}~5d_{3/2}$          &  6 & 6-&(31095.82)& 31189/&$-$0.30 && 32928/&$-$5.89   \\
$4f^{11}~(^4I_{11/2})~5d_{5/2}~(11/2,5/2)$  & $4f^{11}~5d$                  &  7 & 6-&(33191.53)& 34004/&$-$2.45 && 35490/&$-$6.93   \\
$4f^{11}~(^4I_{11/2})~5d_{5/2}~(11/2,5/2)$  & $^4I_{11/2}~5d_{5/2}$         &  8 & 6-&(33875.19)& 34893/&$-$3.01 && 36573/&$-$7.96   \\
$4f^{11}~(^4I_{9/2}~)5d_{5/2}~(9/2,5/2)$    & $4f^{11}~5d$                  &  9 & 6-&(35856.62)& 36933/&$-$3.00 && 38635/&$-$7.75   \\
$4f^{11}~(^4F_{9/2}~)5d_{5/2}~(9/2,5/2)$    & $^4I_{9/2}~5d_{5/2}$          & 10 & 6-&(36570.10)& 37627/&$-$2.89 && 39036/&$-$6.74   \\
$4f^{11}~(^4I_{15/2})~5d_{5/2}~(15/2,5/2)$ & $^4I_{13/2}~5d_{5/2}$         &  3 & 5-&(27870.83)& 28262/&$-$1.40 && 30113/&$-$8.04   \\
$4f^{11}~(^4I_{11/2})~5d_{3/2}~(11/2,3/2)$ & $^4I_{11/2}~5d_{3/2}$         &  4 & 5-&(29995.62)& 30345/&$-$1.16 && 32127/&$-$7.10   \\
$4f^{11}~(^4I_{9/2})~5d_{3/2}~(9/2,3/2)$   & $4f^{11}~5d$                  &  5 & 5-&(31214.52)& 31916/&$-$2.25 && 33507/&$-$7.34   \\
$4f^{11}~(^4F_{9/2})~5d_{3/2}~(9/2,3/2)$   & $4f^{11}~5d$                  &  6 & 5-&(32614.37)& 33040/&$-$1.31 && 34629/&$-$6.18   \\
$4f^{11}~(^4F_{9/2})~5d_{3/2}~(9/2,3/2)$   & $4f^{11}~5d$                  &  7 & 5-&(33704.29)& 34444/&$-$2.20 && 36152/&$-$7.26   \\
$4f^{11}~(^4I_{9/2})~5d_{5/2}~(9/2,5/2)$   & $4f^{11}~5d$                  &  8 & 5-&(36330.81)& 37154/&$-$2.27 && 38817/&$-$6.84   \\
$4f^{11}~(^4F_{9/2})~5d_{5/2}~(9/2,5/2)$   & $4f^{11}~5d$                  &  9 & 5-&(36655.60)& 38287/&$-$4.45 && 39879/&$-$8.79   \\
$4f^{11}~(^4F_{7/2})~5d_{3/2}~(7/2,3/2)$   & $4f^{11}~5d$                  & 10 & 5-&(39265.81)& 41260/&$-$5.08 && 42696/&$-$8.74   \\
$4f^{11}~(^2H^2_{11/2})~5d_{3/2}~(11/2,3/2)$& $4f^{11}~5d$                  & 11 & 5-&(40857.10)& 41726/&$-$2.13 && 43204/&$-$5.75   \\
$4f^{11}~(^4F_{7/2})~5d_{5/2}~(7/2,5/2)$   & $4f^{11}~5d$                  & \colorbox[rgb]{0.75,0.75,0.75}{12} & \colorbox[rgb]{0.75,0.75,0.75}{5-}&(46552.18)& \colorbox[rgb]{0.75,0.75,0.75}{42297/}&\colorbox[rgb]{0.75,0.75,0.75}{9.14}    && 43795/&5.92      \\
$4f^{11}~(^2H^2_{11/2})~5d_{5/2}~(11/2,5/2)$& $4f^{11}~5d$                  & \colorbox[rgb]{0.75,0.75,0.75}{13} & \colorbox[rgb]{0.75,0.75,0.75}{5-}&(48747.15)& \colorbox[rgb]{0.75,0.75,0.75}{44128/}&\colorbox[rgb]{0.75,0.75,0.75}{9.48}    && 45642/&6.37      \\
$4f^{11}~(^4I_{13/2})~5d_{5/2}~(13/2,5/2)$ &$^4I_{13/2}~5d_{5/2}$         &  4 & 8-&(28555.40)& 28325/&0.81    && 30112/&$-$5.45   \\
$4f^{11}~(^4I_{11/2})~5d_{5/2}~(11/2,5/2)$ &$^4I_{11/2}~5d_{5/2}$         &  5 & 8-&(31701.46)& 31400/&0.95    && 33108/&$-$4.44   \\
$4f^{11}~(^4I_{11/2})~5d_{3/2}~(11/2,3/2)$ &$^4I_{11/2}~5d_{3/2}$         &  4 & 7-&(28818.44)& 28722/&0.33    && 30546/&$-$5.99   \\
$4f^{11}~(^4I_{13/2})~5d_{5/2}~(13/2,5/2)$ &$^4I_{13/2}~5d_{5/2}$         &  5 & 7-&(29610.99)& 29454/&0.53    && 31269/&$-$5.60   \\
$4f^{11}~(^4I_{11/2})~5d_{5/2}~(11/2,5/2)$ &$^4I_{11/2}~5d_{5/2}$         &  6 & 7-&(32559.55)& 32825/&$-$0.82 && 34542/&$-$6.09   \\
\colorbox[rgb]{0.75,0.75,0.75}{$4f^{11}~(^4I_{9/2})~5d_{5/2}~(9/2,5/2)$}    &\colorbox[rgb]{0.75,0.75,0.75}{$^4F_{9/2}~5d_{5/2}$}          &  \colorbox[rgb]{0.75,0.75,0.75}{7} & \colorbox[rgb]{0.75,0.75,0.75}{7-}&(36636.87)& \colorbox[rgb]{0.75,0.75,0.75}{34516/}&\colorbox[rgb]{0.75,0.75,0.75}{5.79}    && 36135/&1.37      \\
$4f^{11}~(^4I_{11/2})~5d_{5/2}~(11/2,5/2)$ &$^4I_{11/2}~5d_{3/2}$         &  1 & 3-&(29466.42)& 29374/&0.31    && 31265/&$-$6.10   \\
$4f^{11}~(^4I_{9/2})~5d_{5/2}~(9/2,5/2)$   &$^4I_{9/2}~5d_{5/2}$          &  2 & 3-&(31846.16)& 31869/&$-$0.07 && 33733/&$-$5.92   \\
$4f^{11}~(^4F_{9/2})~5d_{3/2}~(9/2,3/2)$   &$^4F_{9/2}~5d_{3/2}$          &  3 & 3-&(33185.64)& 34565/&$-$4.16 && 36072/&$-$8.70   \\
$4f^{11}~(^4I_{9/2})~5d_{3/2}~(9/2,3/2)$   &$4f^{11}~5d$                  &  4 & 3-&(36167.30)& 37527/&$-$3.76 && 39226/&$-$8.46   \\
$4f^{11}~(^4F_{7/2})~5d_{3/2}~(7/2,3/2)$   &$4f^{11}~5d$                  &  5 & 3-&(37812.87)& 38909/&$-$2.90 && 40386/&$-$6.81   \\
$4f^{11}~(^4F_{7/2})~5d_{3/2}~(7/2,3/2)$   &$4f^{11}~5d$                  &  6 & 3-&(38924.30)& 39652/&$-$1.87 && 41122/&$-$5.65   \\
$4f^{11}~(^4F_{9/2})~5d_{5/2}~(9/2,5/2)$   &$4f^{11}~5d$                  &  7 & 3-&(40407.72)& 41020/&$-$1.51 && 41122/&$-$5.01   \\
$4f^{11}~(^4I_{9/2})~5d_{5/2}~(9/2,5/2)$   &$4f^{11}~5d$                  &  \colorbox[rgb]{0.75,0.75,0.75}{1} & \colorbox[rgb]{0.75,0.75,0.75}{2-}&(38563.97)& \colorbox[rgb]{0.75,0.75,0.75}{32408/}&\colorbox[rgb]{0.75,0.75,0.75}{15.96}   && 34263/&11.15     \\
\hline
\enddata 
\tablecomments{The relative difference compared with NIST/(SE) data is given in percent.}
\end{deluxetable*}

\begin{deluxetable*}{lrrrrr r@{\hskip 0.05pt}r r r@{\hskip 0.05pt}rr}[ht!!]
\tabletypesize{\footnotesize}
\tablecaption{\label{energy_final_2} The Proposed Energy Levels in Comparison with NIST/(SE) Data and Their Relative Difference (Given in Percent).}
\tablehead      {
\multicolumn{1}{c}{JP}&\multicolumn{1}{c}{NIST/(SE)} &\multicolumn{1}{c}{iden. in \cite{Wyart}} & \multicolumn{3}{c}{POS} &
\multicolumn{5}{c}{\textbf{SD 5d ZF$^{MCDHF}$ ($AS_4$)}} &\multicolumn{1}{c}{iden. in present work} \\
}
\startdata
4-&(46937.23)&& 10 &$\rightarrow$& 15 & 41524/&11.53   &$\rightarrow$& 47046/&$-$0.23  \\
5-&(46552.18)&& 12 &$\rightarrow$& 16 & 42297/&9.14    &$\rightarrow$& 47467/&$-$1.96  \\
5-&(48747.15)&& 13 &$\rightarrow$& 17 & 44128/&9.48    &$\rightarrow$& 48836/&$-$0.18  \\
7-&(36636.87)&$^4F_{9/2}~5d_{5/2}$&  7 &$\rightarrow$&  8 & 34516/&5.79    &$\rightarrow$& 37645/&$-$2.75 & $4f^{11}(^4F_{9/2})~5d_{5/2}~(9/2,5/2)$ \\
2-&(38563.97)&&  1 &$\rightarrow$&  3 & 32408/&15.96   &$\rightarrow$& 39173/&$-$1.58  \\		
\hline
\enddata 
\end{deluxetable*}

The full energy spectrum (energy levels for 399 states) with unique labels  
and with atomic state function composition in $LS$ coupling using the \textbf{SD 5d ZF$^{MCDHF}$} strategy is presented
in machine-readable format in Table \ref{energies}.

\begin{deluxetable*}{rrrrrll}[ht!!]
	\tabletypesize{\footnotesize}
	\setlength{\tabcolsep}{10pt}
	\tablecaption{
Energy Levels (in cm$^{-1}$) and Atomic State Function Composition of the Ground 
[Xe]$4f^{12}$ and First Excited [Xe]$4f^{11}5d$ Configurations for the Er$^{2+}$ ion. \label{energies}}
	\tablehead{ \colhead{No.}& \colhead{POS}  & \colhead{$J$} & \colhead{P} & \colhead{$E$} & \colhead{label} & \colhead{comp.}  }
	\startdata
  1&  1&   6&   +  &      0.00&  $4f^{12}~^3H$                &  0.95                                                                     \\
  2&  1&   4&   +  &   5730.98&  $4f^{12}~^3F$                &  0.56   + 0.29 $4f^{12}~^1G$ + 0.11 $4f^{12}~^3H$                         \\
  3&  1&   5&   +  &   6813.74&  $4f^{12}~^3H$                &  0.95                                                                     \\
  4&  2&   4&   +  &  10889.64&  $4f^{12}~^3H$                &  0.59   + 0.29 $4f^{12}~^3F$ + 0.07 $4f^{12}~^1G$                         \\
  5&  1&   3&   +  &  13267.28&  $4f^{12}~^3F$                &  0.95                                                                     \\
  6&  1&   2&   +  &  14417.89&  $4f^{12}~^3F$                &  0.81   + 0.13 $4f^{12}~^1D$                                              \\
  7&  1&   6&  $-$ &  16391.49&  $4f^{11}~(^{4}I^{1})~5d~^5G$ &  0.75   + 0.13 $4f^{11}~(^4I^1)~5d~^5H$ + 0.02 $4f^{11}~(^{2}K^1)~5d~^3H$ \\
  8&  1&   7&  $-$ &  17333.10&  $4f^{11}~(^{4}I^{1})~5d~^5H$ &  0.76   + 0.11 $4f^{11}~(^4I^1)~5d~^5I$ + 0.02 $4f^{11}~(^{4}G^1)~5d~^5H$ \\
  9&  3&   4&   +  &  18392.93&  $4f^{12}~^1G$                &  0.59   + 0.25 $4f^{12}~^3H$            + 0.11 $4f^{12}~^3F$              \\    
 10&  1&   9&  $-$ &  18523.58&  $4f^{11}~(^{4}I^{1})~5d~^3L$ &  0.43   + 0.33 $4f^{11}~(^4I^1)~5d~^5L$ + 0.17 $4f^{11}~(^{4}I^1)~5d~^5K$ \\
 11&  1&  10&  $-$ &  19537.18&  $4f^{11}~(^{4}I^{1})~5d~^5L$ &  0.92   + 0.02 $4f^{11}~(^2K^1)~5d~^3M$                                   \\
 12&  1&   8&  $-$ &  19671.88&  $4f^{11}~(^{4}I^{1})~5d~^5I$ &  0.44   + 0.28 $4f^{11}~(^4I^1)~5d~^5K$ + 0.15 $4f^{11}~(^{4}I^1)~5d~^3K$ \\
 13&  2&   9&  $-$ &  21341.61&  $4f^{11}~(^{4}I^{1})~5d~^5K$ &  0.71   + 0.21 $4f^{11}~(^4I^1)~5d~^3L$                                   \\
 14&  1&   5&  $-$ &  21604.17&  $4f^{11}~(^{4}I^{1})~5d~^5G$ &  0.65   + 0.17 $4f^{11}~(^4I^1)~5d~^5H$ + 0.05 $4f^{11}~(^{4}I^1)~5d~^3G$ \\
 15&  2&   6&  $-$ &  22420.54&  $4f^{11}~(^{4}I^{1})~5d~^5H$ &  0.54   + 0.16 $4f^{11}~(^4I^1)~5d~^5G$ + 0.11 $4f^{11}~(^{4}I^1)~5d~^5I$ \\
 16&  2&   8&  $-$ &  22628.77&  $4f^{11}~(^{4}I^{1})~5d~^3K$ &  0.29   + 0.40 $4f^{11}~(^4I^1)~5d~^5I$ + 0.09 $4f^{11}~(^{4}I^1)~5d~^5L$ \\
 17&  2&   7&  $-$ &  23392.31&  $4f^{11}~(^{4}I^{1})~5d~^3I$ &  0.23   + 0.31 $4f^{11}~(^4I^1)~5d~^5I$ + 0.17 $4f^{11}~(^{4}I^1)~5d~^5K$ \\
 18&  3&   8&  $-$ &  25069.42&  $4f^{11}~(^{4}I^{1})~5d~^5L$ &  0.49   + 0.25 $4f^{11}~(^4I^1)~5d~^3K$ + 0.20 $4f^{11}~(^{4}I^1)~5d~^3L$ \\
 19&  1&   4&  $-$ &  26268.28&  $4f^{11}~(^{4}I^{1})~5d~^5G$ &  0.59   + 0.25 $4f^{11}~(^4I^1)~5d~^5H$ + 0.03 $4f^{11}~(^{2}H^2)~5d~^3F$ \\
 20&  2&   5&  $-$ &  26463.59&  $4f^{11}~(^{4}I^{1})~5d~^5H$ &  0.45   + 0.24 $4f^{11}~(^4I^1)~5d~^3G$ + 0.09 $4f^{11}~(^{4}I^1)~5d~^5I$ \\
			\enddata
\tablecomments{ Table~\ref{energies} is published in its entirety in the machine-readable format. Part of the values are shown here for guidance regarding its form and content.}
\end{deluxetable*}

\section{Transition data results}
The wave functions from the \textbf{SD 5d} and \textbf{SD 5d ZF$^{MCDHF}$} strategies, which were chosen as the optimal computational schemes,
were used to compute E1 transition data between states of the [Xe]$4f^{12}$ and [Xe]$4f^{11}5d$ configurations.
The accuracy of the transition data obtained in this work was evaluated by:
\begin{enumerate}
	\item calculating parameter $dT$, which shows the disagreement between the length and velocity forms of the computed transition rates;
	\item analyzing the convergence of the computed transition rates in the length and velocity forms;
	\item analyzing the dependence of the transition rate on the gauge parameter $G$;
	\item analyzing the dependence of cancellation factor on the gauge parameter $G$;
	\item comparing computed transition data with other experimental or theoretical calculations.
\end{enumerate}

For these investigations a few strong transitions have been chosen as examples. 
The evaluation of transition data will be presented in the sections below.

Computed transition data, such as wavelengths, weighted oscillator strengths, transition rates of E1 along with the accuracy indicator $dT$, are given in machine-readable format in Table \ref{transition_data}.

\subsection{Disagreement between the length and velocity and their convergence}
In a variational approach the wave functions are optimized on an energy expression.
In general this gives a better representation of the outer part of the wave functions, thus favoring the length form.
The velocity form contains a dependence on the transition energy in the matrix element, which may affect the accuracy 
of the evaluation. Due to the above mentioned reasons, a much slower convergence of the velocity gauge is expected \citep{Fischer_1995}.
However, a recent paper by \citet{Asimina_2019}, analyzing in detail the convergence properties of transitions in light elements, suggests that  transition probabilities in the
Coulomb gauge may give the more accurate values. Thus, it is important to systematically study the 
transition data to see which gauge results in the most rapid convergence.

\begin{figure}[ht!]
\includegraphics[width=0.47\textwidth]{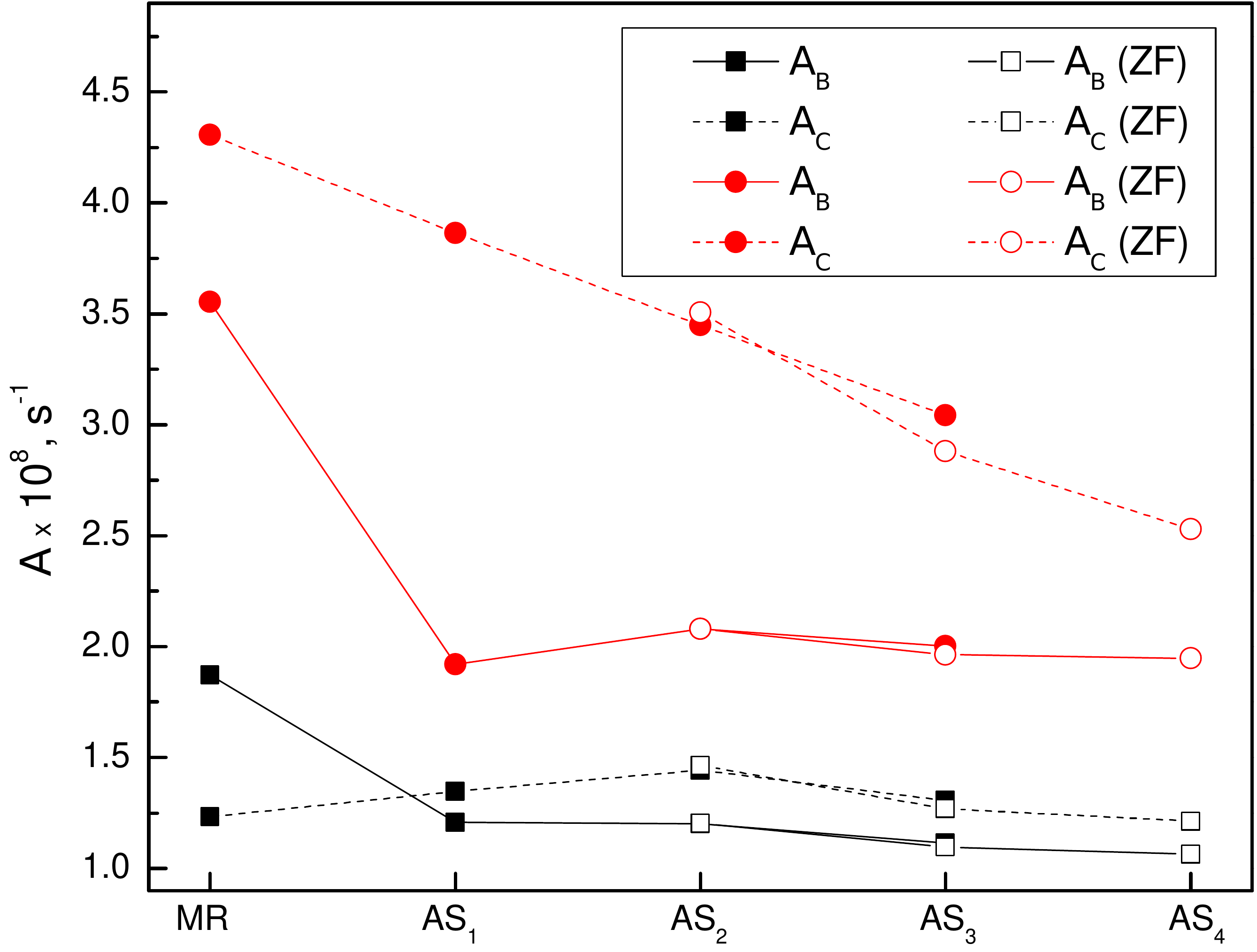}
\caption{Convergence of E1 transition probabilities using the \textbf{SD 5d} strategy 
(open symbols mark the results when the \textbf{ZF$^{MCDHF}$} approach is applied).
The $4f^{12}~{}^3P_0$ -- $4f^{11}~{}(^2F^1)~{}5d~{}^3P_1$ transition is marked in black and the $4f^{12}~{}^1S_0$ -- $4f^{11}~{}(^2F^1)~{}5d~{}^1P_1$ transition in red.}
\label{konvergencija_suoliu}
\end{figure}

\begin{figure}[ht!]
\includegraphics[width=0.47\textwidth]{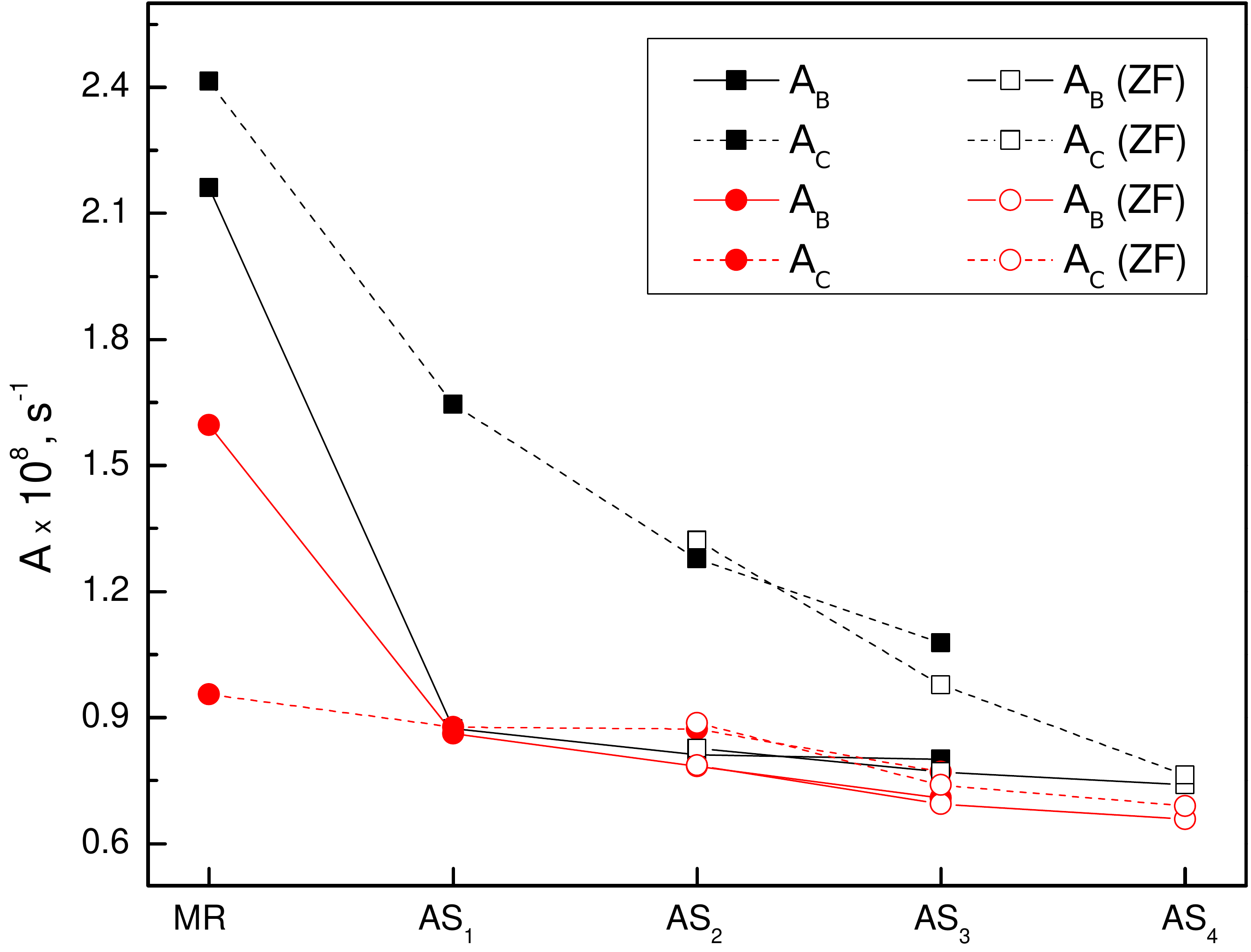}
\caption{Convergence of E1 transition probabilities using the \textbf{SD 5d} strategy 
(open symbols mark the results when the \textbf{ZF$^{MCDHF}$} approach is applied).
The $4f^{12}~{}^3P_2$ -- $4f^{11}~{}(^2F^2)~{}5d~{}^1P_1$ transition is marked in black and the $4f^{12}~{}^3P_2$ -- $4f^{11}~{}(^2F^1)~{}5d~{}^3P_1$ transition in red.}
\label{konvergencija_suoliu_2}
\end{figure}

The convergence of the transition rates in both gauges with the increasing active spaces is presented 
in Figures \ref{konvergencija_suoliu} and \ref{konvergencija_suoliu_2}.
From these Figures it is seen that  transition probabilities in the Babushkin gauge are more stable to electron correlation 
effects than the probabilities in the Coulomb gauge.
The $dT$ for the analyzed transitions based on the final $AS_4$ in the \textbf{SD 5d ZF$^{MCDHF}$} strategy are
12\% for $4f^{12}~{}^3P_0$ -- $4f^{11}~{}(^2F^1)~{}5d~{}^3P_1$, 
23\% for $4f^{12}~{}^1S_0$ -- $4f^{11}~{}(^2F^1)~{}5d~{}^1P_1$ (Figure \ref{konvergencija_suoliu}); 
3\% for $4f^{12}~{}^3P_2$ -- $4f^{11}~{}(^2F^2)~{}5d~{}^1P_1$ and 5\% for$4f^{12}~{}^3P_2$ -- $4f^{11}~{}(^2F^1)~{}5d~{}^3P_1$(Figure \ref{konvergencija_suoliu_2}).

Analyzing the impact of the ZF method on the transition rates, we see that ZF$^{MCDHF}$ $AS_3$ 
 reduces transition rates compared to those from the \textbf{SD 5d} strategy. The transition rates in Coulomb gauge change even more than those in 
the Babushkin gauge.
Transition rates in Babushkin gauge decreases just by a few percent for the analyzed transitions.
The above analysis shows that the Babushkin gauge is the preferred one.

\subsection{Gauge dependence}
\begin{figure}[ht!]
\includegraphics[width=0.47\textwidth]{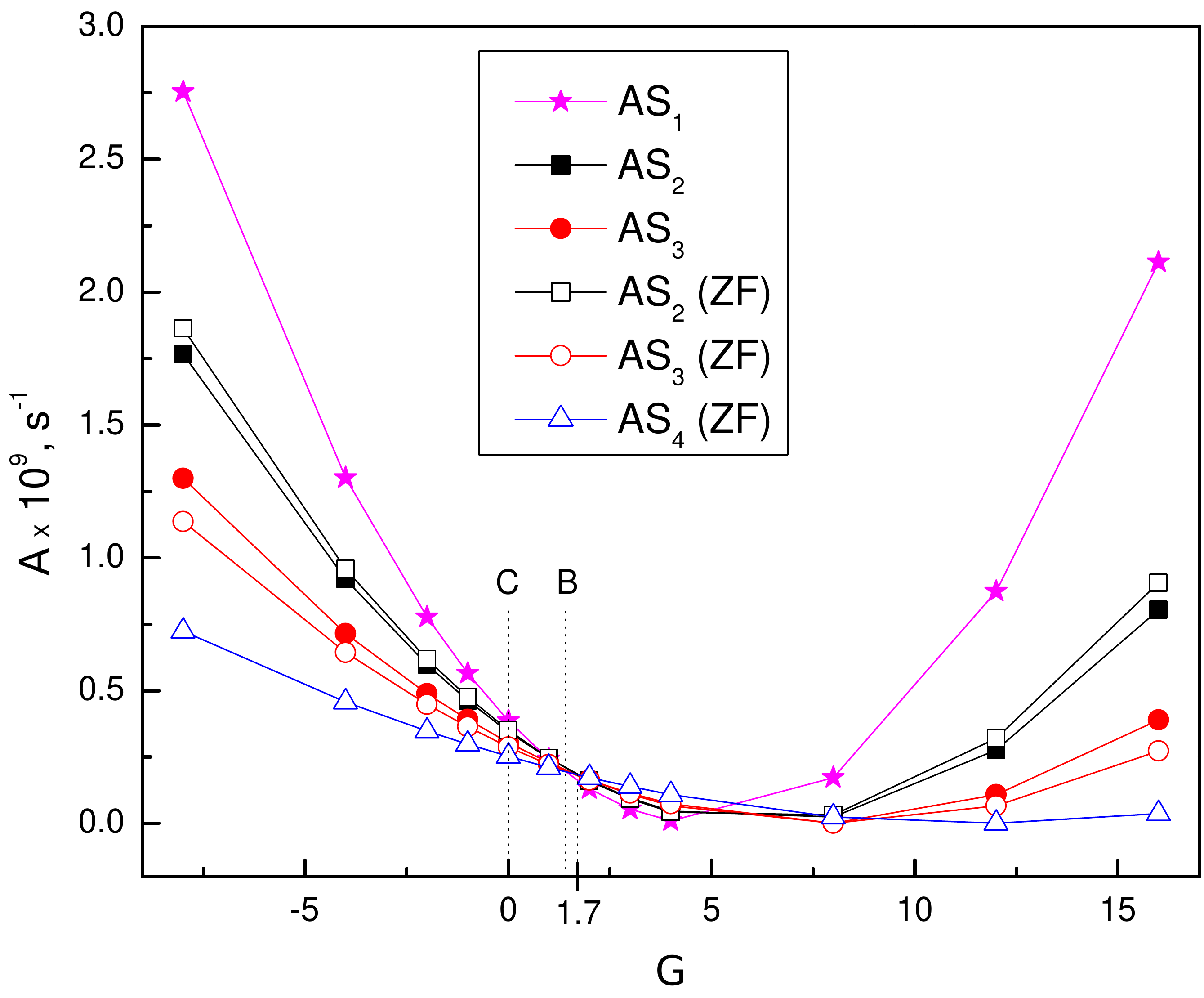}
\caption{The gauge dependence of the $4f^{12}~{}^1S_0$ -- $4f^{11}~{}(^2F^1)~{}5d~{}^1P_1$ E1 transition probability for the
different active space calculations using the \textbf{SD 5d} strategy (open symbols mark the results when the \textbf{ZF$^{MCDHF}$} approach is applied).}
\label{G_dependence_1S_1P_plus_ZF}
\end{figure}

\begin{figure}[ht!]
\includegraphics[width=0.47\textwidth]{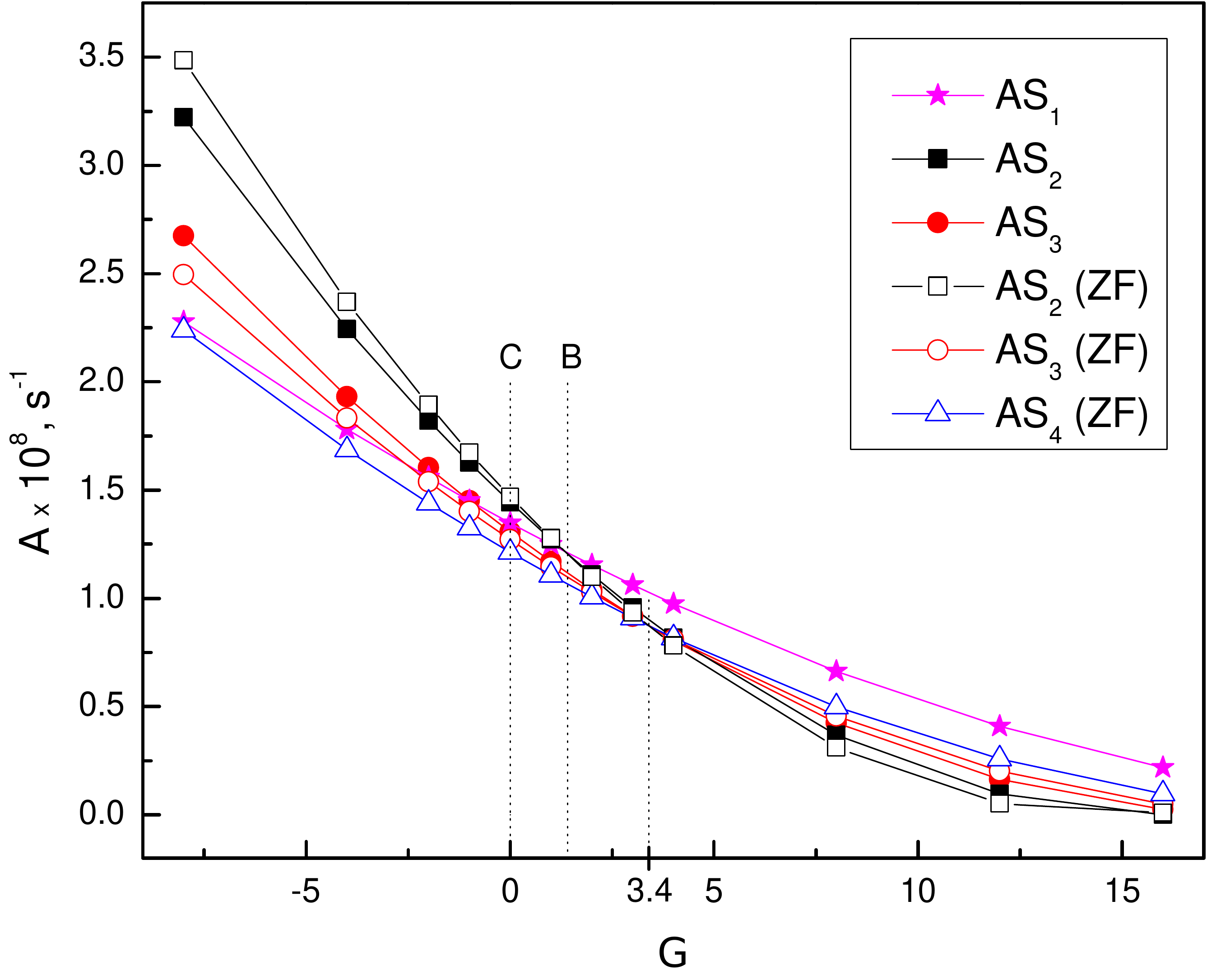}
\caption{The gauge dependence of the $4f^{12}~{}^3P_0$ -- $4f^{11}~{}(^2F^1)~{}5d~{}^3P_1$ E1 transition probability for the
different active space calculations using the \textbf{SD 5d} strategy (open symbols mark the results when the \textbf{ZF$^{MCDHF}$} approach is applied).}
\label{G_dependence_3P_3P_plus_ZF}
\end{figure}

\begin{figure}[ht!]
\includegraphics[width=0.47\textwidth]{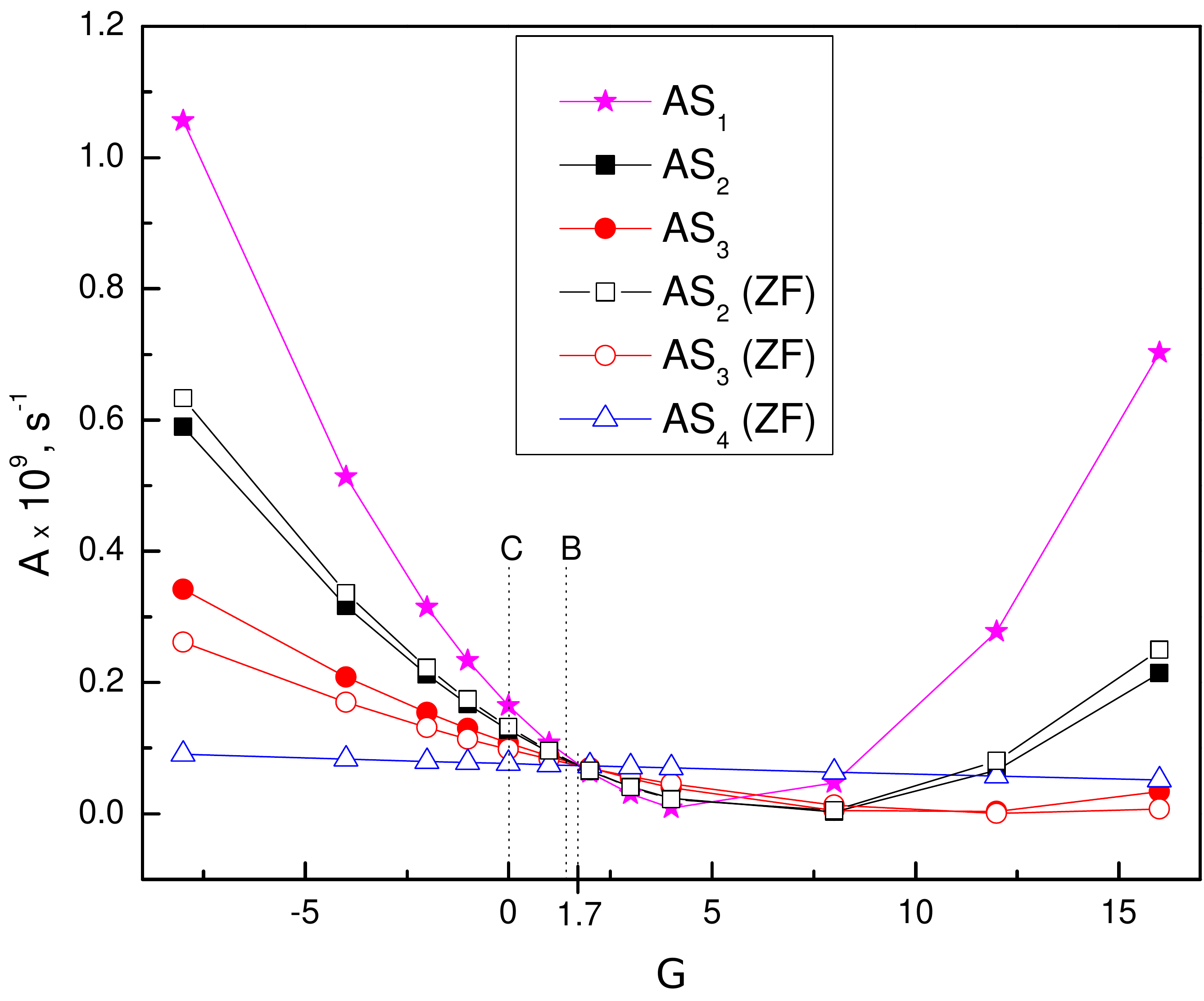}
\caption{The gauge dependence of the $4f^{12}~{}^3P_2$ -- $4f^{11}~{}(^2F^2)~{}5d~{}^1P_1$ E1 transition probability for the
different active space calculations using the \textbf{SD 5d} strategy (open symbols mark the results when the \textbf{ZF$^{MCDHF}$} approach is applied).}
\label{G_dependence_2_3P_1P_be_MR_plus_ZF}
\end{figure}

\begin{figure}[ht!]
\includegraphics[width=0.47\textwidth]{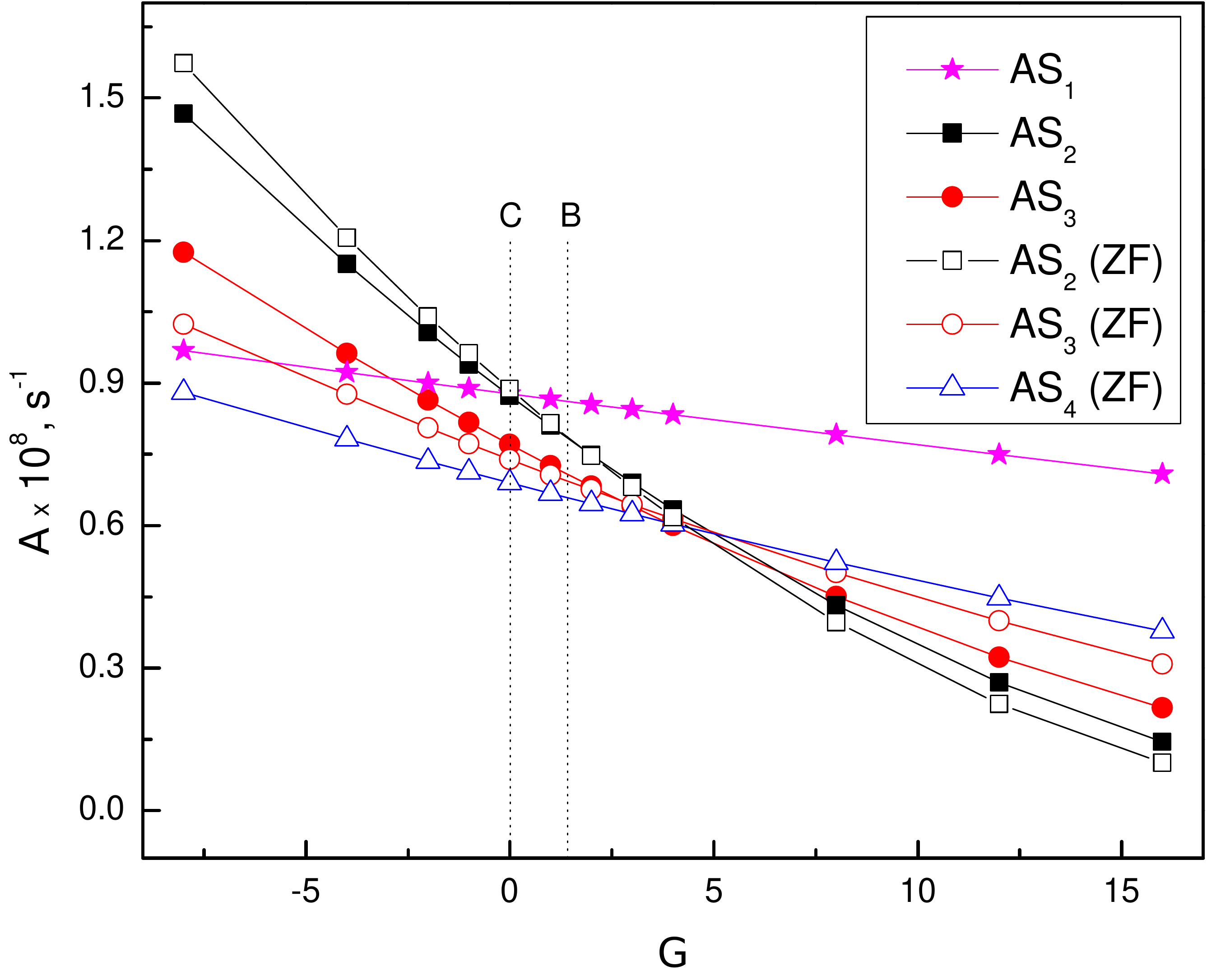}
\caption{The gauge dependence of the $4f^{12}~{}^3P_2$ -- $4f^{11}~{}(^2F^1)~{}5d~{}^3P_1$ E1 transition probability for the
different active space calculations using the \textbf{SD 5d} strategy (open symbols mark the results when the \textbf{ZF$^{MCDHF}$} approach is applied).}
\label{G_dependence_2_3P_3P_be_MR_plus_ZF}
\end{figure}

In Figures \ref{G_dependence_1S_1P_plus_ZF}--\ref{G_dependence_2_3P_3P_be_MR_plus_ZF} the dependence of the transition
probabilities for the different active space calculations on the gauge parameter $G$ is displayed. In each of these Figures the position of Coulomb and Babushkin gauges are marked by dotted lines. 
For some of analyzed transitions the curves of gauge dependence intersect at some point. The cross points are marked by dotted lines and the values are placed on the axis.
The curves cross at around $G=1.7$ (very close to the Babushkin form) 
for the $4f^{12}~{}^1S_0$ -- $4f^{11}~{}(^2F^1)~{}5d~{}^1P_1$ (Figure \ref{G_dependence_1S_1P_plus_ZF}) 
and $4f^{12}~{}^3P_2$ -- $4f^{11}~{}(^2F^1)~{}5d~{}^1P_1$ (Figure \ref{G_dependence_2_3P_1P_be_MR_plus_ZF}) transitions.
For the $4f^{12}~{}^3P_0$ -- $4f^{11}~{}(^2F^1)~{}5d~{}^3P_1$ transition (Figure \ref{G_dependence_3P_3P_plus_ZF}) the most of curves (except the curve of gauge dependence with $AS_1$) intersect at around $G=3.4$. 
In case of the $4f^{12}~{}^3P_2$ -- $4f^{11}~{}(^2F^1)~{}5d~{}^3P_1$ transition (Figure \ref{G_dependence_2_3P_3P_be_MR_plus_ZF}) the curves do not intersect at one point.
From these Figures we can see that by increasing the active space, the curves of gauge dependence approach straight lines.
At $AS_4$ (final results) these curves are very close to straight lines. It means that the wave functions should be quite accurate.

\subsection{Cancellation factor}
\begin{figure}[ht!]
\includegraphics[width=0.47\textwidth]{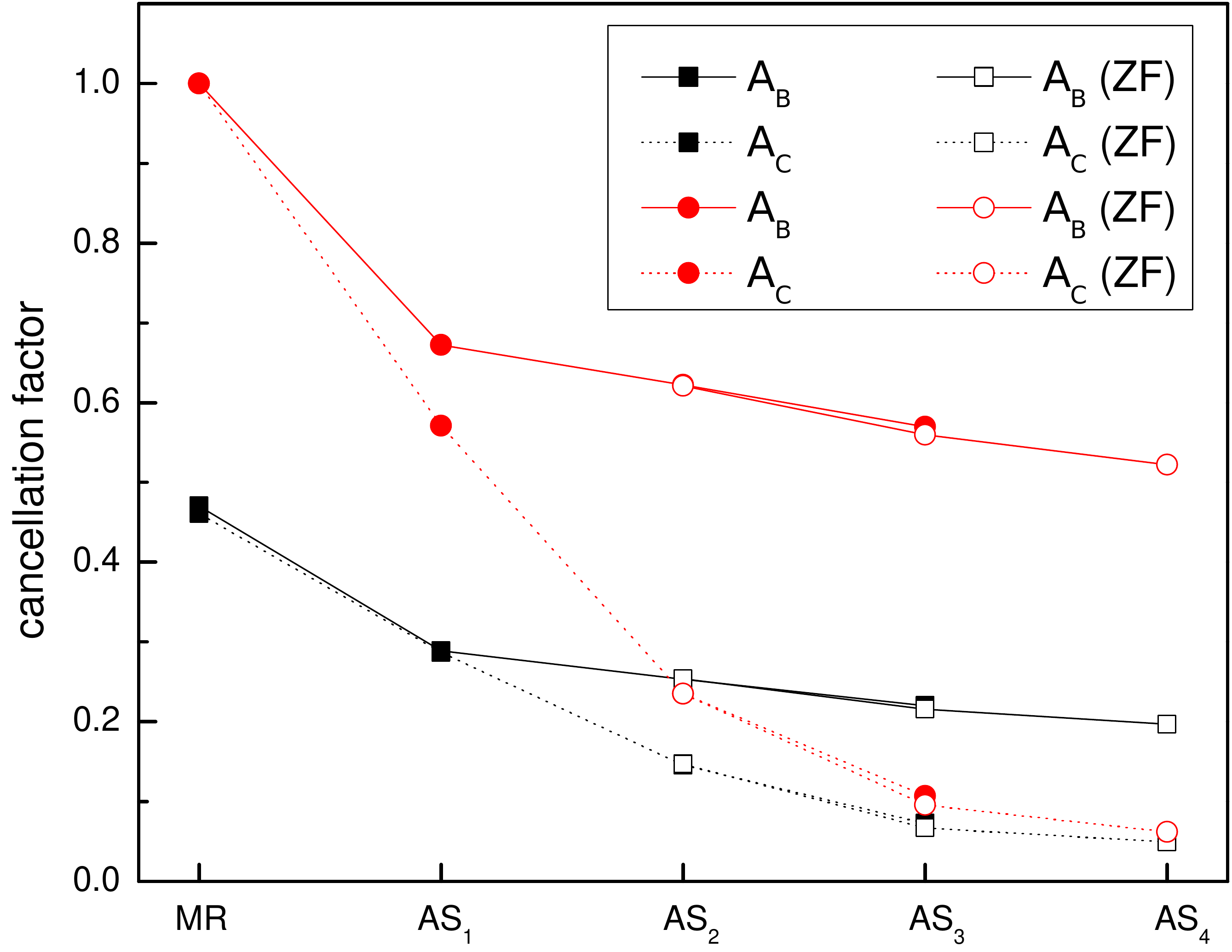}
\caption{Cancellation factor dependence on the active space.
The $4f^{12}~{}^3P_0$ -- $4f^{11}~{}(^2F^1)~{}5d~{}^3P_1$ transition is marked in black and the $4f^{12}~{}^1S_0$ -- $4f^{11}~{}(^2F^1)~{}5d~{}^1P_1$ transition in red.}
\label{CF_nuo_AS}
\end{figure}

\begin{figure}[ht!]
\includegraphics[width=0.47\textwidth]{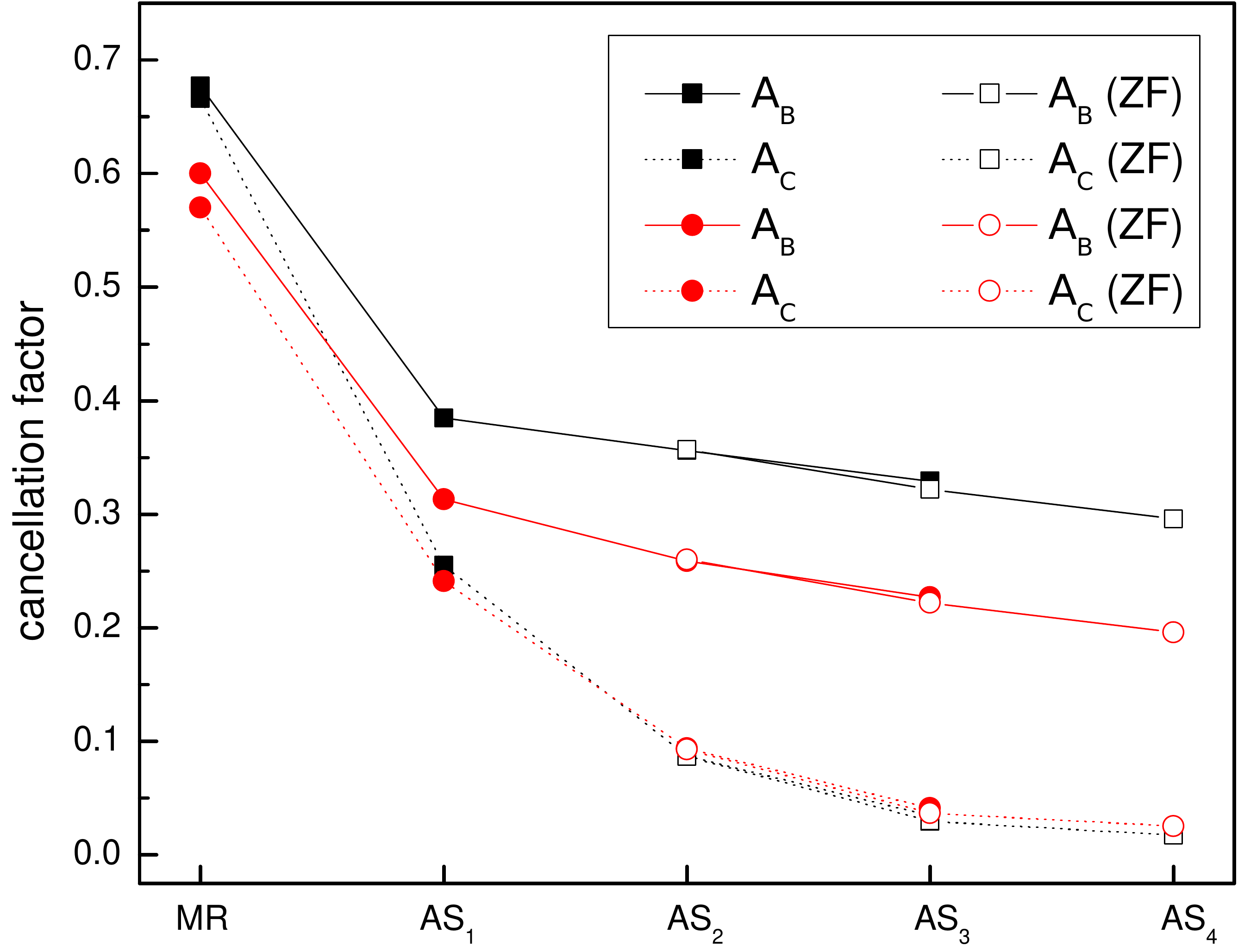}
\caption{Cancellation factor dependence on the active space.
The $4f^{12}~{}^3P_2$ -- $4f^{11}~{}(^2F^2)~{}5d~{}^1P_1$ transition is marked in black and the $4f^{12}~{}^3P_2$ -- $4f^{11}~{}(^2F^1)~{}5d~{}^3P_1$ transition in red.}
\label{CF_nuo_AS__2}
\end{figure}

\begin{figure}[ht!]
\includegraphics[width=0.47\textwidth]{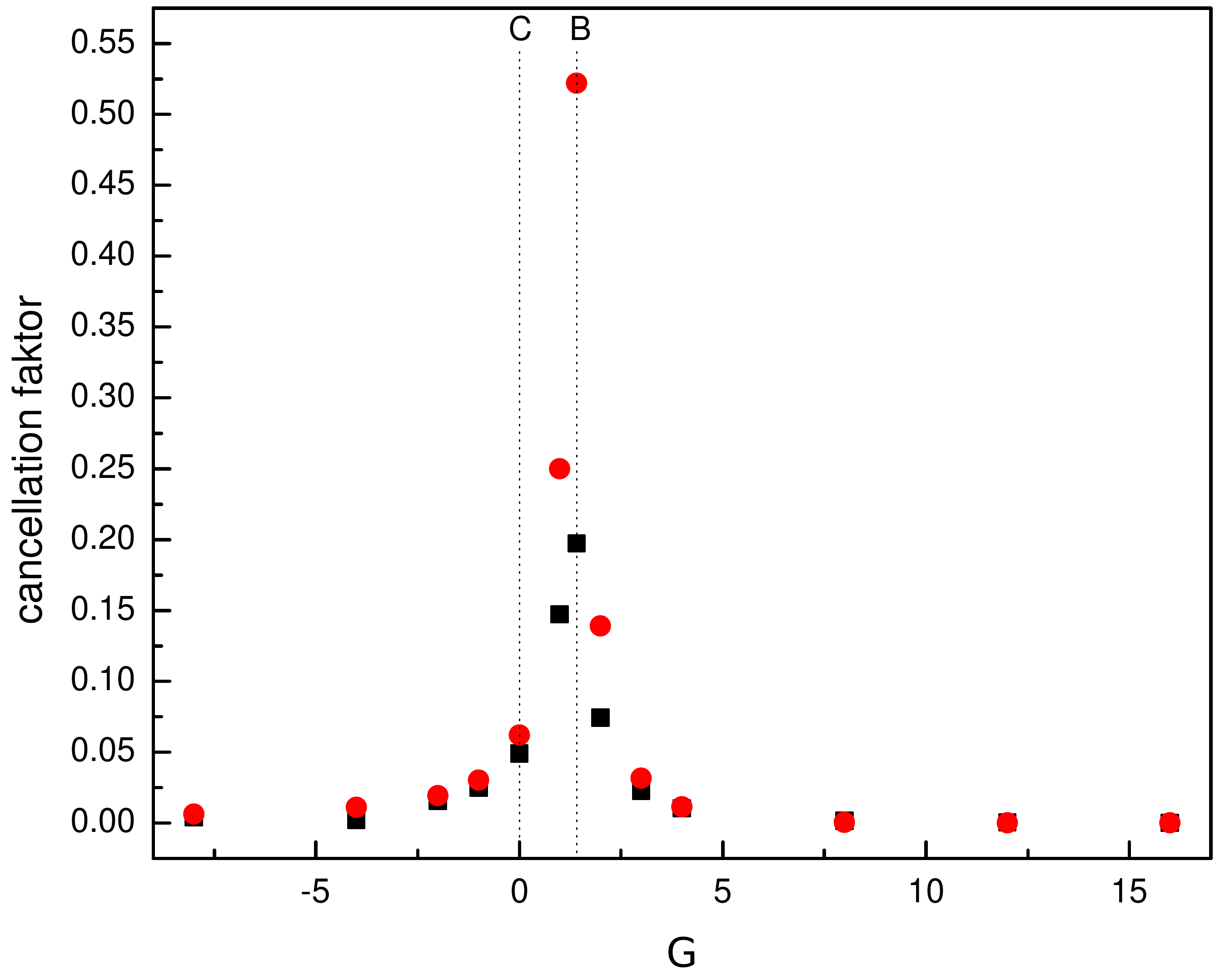}
\caption{Cancellation factor dependence on gauge using the \textbf{SD 5d ZF$^{MCDHF}$} strategy (at $AS_4$).
The $4f^{12}~{}^3P_0$ -- $4f^{11}~{}(^2F^1)~{}5d~{}^3P_1$ transition is marked and in black and the
$4f^{12}~{}^1S_0$ -- $4f^{11}~{}(^2F^1)~{}5d~{}^1P_1$ transition is marked in red.}
\label{cancellation_factor}
\end{figure}

\begin{figure}[ht!]
\includegraphics[width=0.47\textwidth]{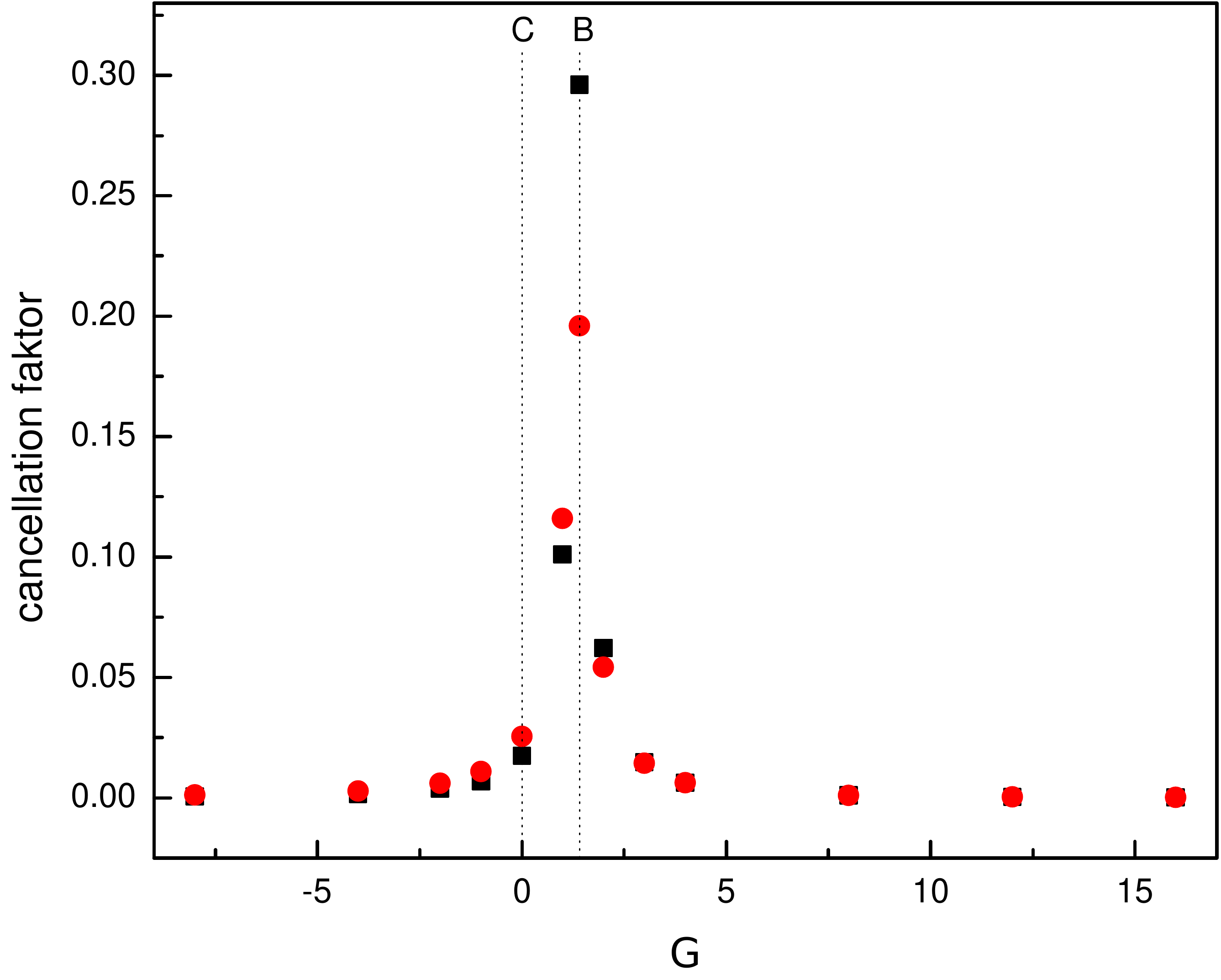}
\caption{Cancellation factor dependence on gauge using the \textbf{SD 5d ZF$^{MCDHF}$} strategy (at $AS_4$).
The $4f^{12}~{}^3P_2$ -- $4f^{11}~{}(^2F^2)~{}5d~{}^1P_1$ transition is marked and in black and
the $4f^{12}~{}^3P_2$ -- $4f^{11}~{}(^2F^1)~{}5d~{}^3P_1$ transition is marked in red.}
\label{cancellation_factor_2}
\end{figure}

Figures \ref{CF_nuo_AS} and \ref{CF_nuo_AS__2} show the CF as a function of the increasing active space
for the \textbf{SD 5d} strategy. From the Figures it is seen that CF in the Babushkin gauge for the analyzed transitions in all active spaces are lager than in the Coulomb gauge.
In Figure \ref{cancellation_factor} and \ref{cancellation_factor_2} we present the dependence of CF on the gauge parameter $G$ 
using the \textbf{SD 5d ZF}$^{MCDHF}$ strategy (at $AS_4$). The CF is presented for the four analyzed transitions.
The CFs in Babushkin gauge for these transitions are much larger than 0.1 or 0.05, and in all cases they are the largest ones. They are even larger than at the 
cross points, where gauge dependence curves from different active spaces intersect.
The CFs in Coulomb gauge for the transitions
$4f^{12}~{}^3P_2$ -- $4f^{11}~{}(^2F^2)~{}5d~{}^1P_1$ and $4f^{12}~{}^3P_2$ -- $4f^{11}~{}(^2F^1)~{}5d~{}^3P_1$ (Figure \ref{cancellation_factor_2}) are smaller 
than 0.05, which  means that in velocity form there is a strong cancellation effect.
For the $4f^{12}~{}^3P_0$ -- $4f^{11}~{}(^2F^1)~{}5d~{}^3P_1$ and $4f^{12}~{}^1S_0$ -- $4f^{11}~{}(^2F^1)~{}5d~{}^1P_1$ 
(Figure \ref{cancellation_factor}) transitions, the CF in the Coulomb gauge is around 0.05.
The analysis shows that transition data in the Babushkin gauge are less affected by cancellation effects than transition data in the velocity gauge.

\subsection{Comparison with other computations}
No experimental transition rates for the studied configurations of Er$^{2+}$ are available.
The transition data obtained using the \textbf{SD 5d ZF$^{MCDHF}$} strategy (at $AS_4$)
are compared with rates presented by \citet{Wyart} and \citet{Biemont}. They used experimental transition wavelengths to compute transition data. 
\citet{Biemont} used the Cowan code and included core-polarization effects in the computations.

\begin{figure}[ht!]
\includegraphics[width=0.47\textwidth]{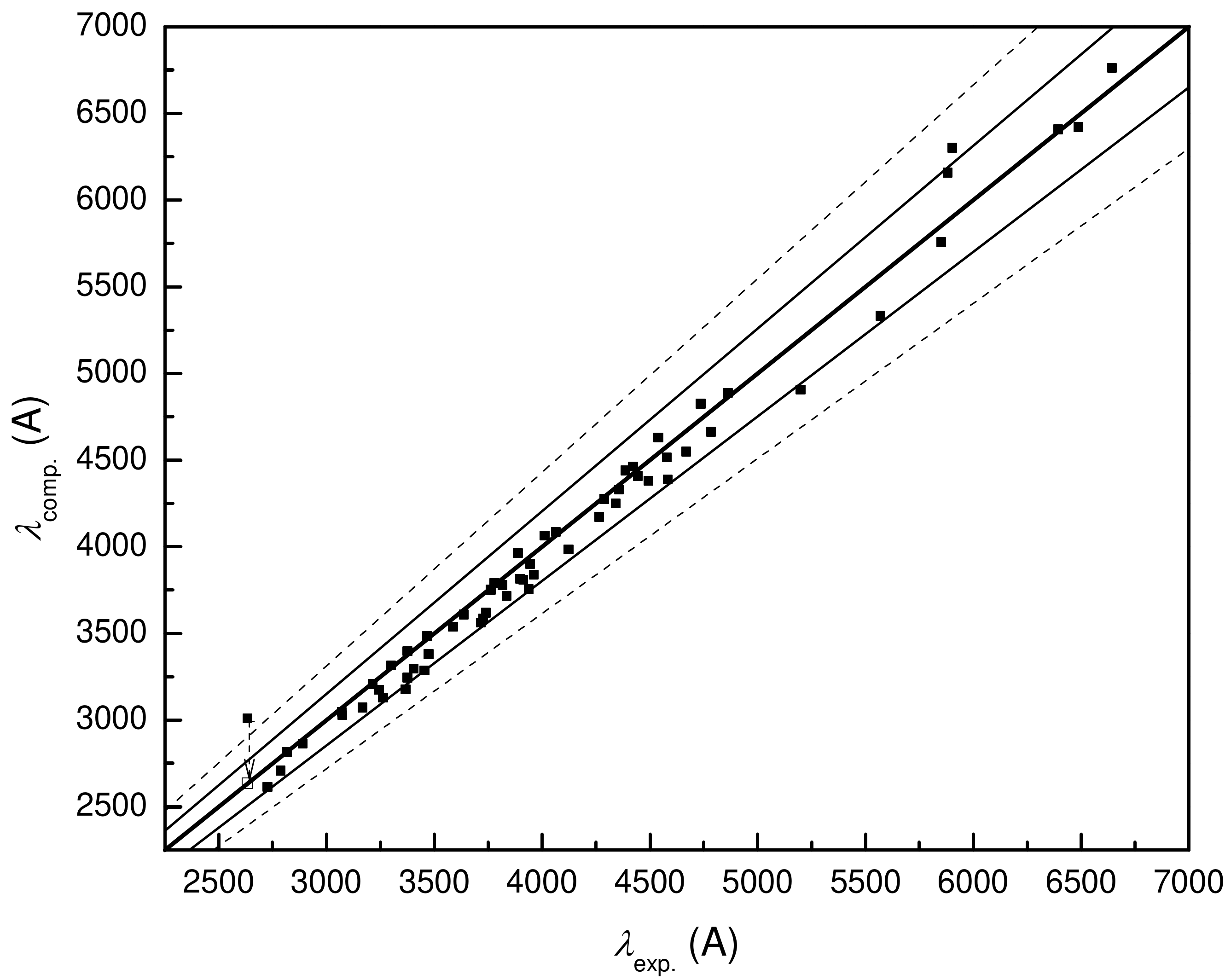}
\caption{Comparison of transition wavelengths between our computed data (comp.) 
using the \textbf{SD 5d ZF$^{MCDHF}$} strategy (at $AS_4$) and experimental data presented in the paper by Wyart et al. \cite{Wyart}. The thick line corresponds to perfect agreement, while thin solid and dashed lines
correspond to 5\% and 10\% deviations. The dashed arrows indicate the improved agreement by applying replacement in the energy levels discussed in Section \ref{section_final_results}.}
\label{comp_wavelengths}
\end{figure}

\begin{figure}[ht!]
\includegraphics[width=0.47\textwidth]{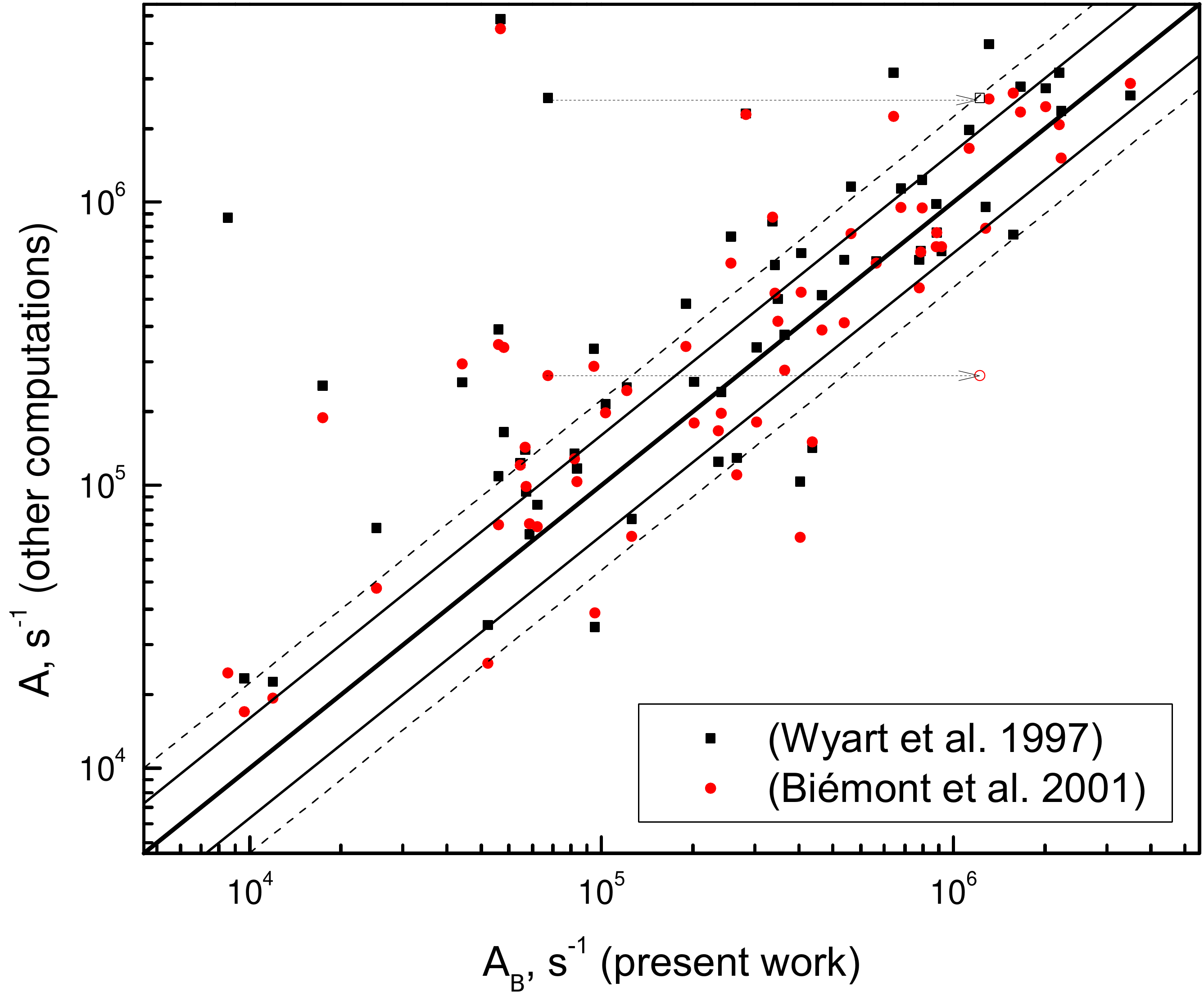}
\caption{Comparison of transition rates of present work ($A$ is given in Babushkin gauge) with rates presented in \cite{Wyart} and \cite{Biemont}. 
The data from \cite{Wyart} are marked by black squares and the red circles correspond to the results by Bi\'emont et al. \cite{Biemont}.
The thick line corresponds to perfect agreement, while the thin solid and dashed lines 
correspond to deviations by factors of 1.5 and 2.0, respectively. The dashed arrows indicate the improved agreement by applying replacement in the energy levels discussed in Section \ref{section_final_results}.}
\label{comp_with_wyart}
\end{figure}

Figure \ref{comp_wavelengths} presents a comparison of obtained transition wavelengths with experimental data, 
which were presented in the paper by \citet{Wyart}. The agreement between the computed wavelengths 
and the experimental ones is very good. Almost all compared lines achieve 5\% uncertainty. 
In Figure \ref{comp_with_wyart} the comparison of transition rates (given in Babushkin gauge) of the present work with rates 
available from other computations \citep{Wyart,Biemont} is displayed. 
It is seen that there is a good agreement with values from other authors for the stronger transitions. 
However, the transitions presented in the Figure are not the strongest obtained in this work. 
The strongest transition have rates of the order $10^8$ s$^{-1}$.
By applying replacement in the energy levels discussed in Section \ref{section_final_results} we achieve 
better agreement for wavelength and transition rate of marked transition
(see open symbols in Figures \ref{comp_wavelengths} and \ref{comp_with_wyart}).

\begin{deluxetable*}{rrrlrrrlrrrrr}[ht!!]
	\tabletypesize{\tiny}
	\setlength{\tabcolsep}{10pt}
	\tablecaption{
Transition Energies $\Delta E$ (in cm$^{-1}$), Transition Wavelengths $\lambda$ (in \AA), 
Weighted Oscillator Strengths $gf$ and Transition Rates $A$ (in s$^{-1}$) for E1 Transitions of the Er$^{2+}$ Ion. \label{transition_data}}
	\tablehead{ \colhead{No.(u)}& \colhead{$J_u$}  & \colhead{P$_u$} & \colhead{state(u)} & \colhead{No.(l)}& \colhead{$J_l$}  & \colhead{P$_l$} & \colhead{state(l)} & \colhead{$\Delta E$} & \colhead{$\lambda$} & \colhead{$A$} & \colhead{$gf$} & \colhead{$dT$} }
	\startdata
  7&   6& $-$& $4f^{11}~(^4I^1)~5d~^5G$ &      1&   6& +& $4f^{12}~^3H$            &    16391&     6100.73&  8.417E+02&  6.106E$-$05&  0.855 \\
  7&   6& $-$& $4f^{11}~(^4I^1)~5d~^5G$ &      3&   5& +& $4f^{12}~^3H$            &     9577&    10440.87&  1.402E+01&  2.978E$-$06&  0.919 \\
  8&   7& $-$& $4f^{11}~(^4I^1)~5d~^5H$ &      1&   6& +& $4f^{12}~^3H$            &    17333&     5769.31&  4.336E$-$01&  3.246E$-$08&  1.000 \\
 14&   5& $-$& $4f^{11}~(^4I^1)~5d~^5G$ &      1&   6& +& $4f^{12}~^3H$            &    21604&     4628.74&  5.100E+04&  1.802E$-$03&  0.711 \\
 14&   5& $-$& $4f^{11}~(^4I^1)~5d~^5G$ &      2&   4& +& $4f^{12}~^3F$            &    15873&     6299.93&  6.063E+04&  3.968E$-$03&  0.685 \\
 14&   5& $-$& $4f^{11}~(^4I^1)~5d~^5G$ &      3&   5& +& $4f^{12}~^3H$            &    14790&     6761.13&  9.658E+03&  7.280E$-$04&  0.771 \\
 14&   5& $-$& $4f^{11}~(^4I^1)~5d~^5G$ &      4&   4& +& $4f^{12}~^3H$            &    10714&     9333.12&  4.760E+03&  6.837E$-$04&  0.780 \\
 14&   5& $-$& $4f^{11}~(^4I^1)~5d~^5G$ &      9&   4& +& $4f^{12}~^1G$            &     3211&    31140.62&  4.578E+01&  7.321E$-$05&  0.962 \\
 15&   6& $-$& $4f^{11}~(^4I^1)~5d~^5H$ &      1&   6& +& $4f^{12}~^3H$            &    22420&     4460.20&  3.113E+05&  1.207E$-$02&  0.551 \\
 15&   6& $-$& $4f^{11}~(^4I^1)~5d~^5H$ &      3&   5& +& $4f^{12}~^3H$            &    15606&     6407.46&  1.165E+04&  9.318E$-$04&  0.700 \\
 17&   7& $-$& $4f^{11}~(^4I^1)~5d~^3I$ &      1&   6& +& $4f^{12}~^3H$            &    23392&     4274.91&  7.104E+05&  2.920E$-$02&  0.517 \\
 19&   4& $-$& $4f^{11}~(^4I^1)~5d~^5G$ &      2&   4& +& $4f^{12}~^3F$            &    20537&     4869.19&  8.829E+02&  2.824E$-$05&  0.084 \\
 19&   4& $-$& $4f^{11}~(^4I^1)~5d~^5G$ &      3&   5& +& $4f^{12}~^3H$            &    19454&     5140.19&  9.666E+03&  3.446E$-$04&  0.595 \\
 19&   4& $-$& $4f^{11}~(^4I^1)~5d~^5G$ &      4&   4& +& $4f^{12}~^3H$            &    15378&     6502.53&  2.340E+03&  1.335E$-$04&  0.796 \\
 19&   4& $-$& $4f^{11}~(^4I^1)~5d~^5G$ &      5&   3& +& $4f^{12}~^3F$            &    13001&     7691.72&  1.108E+04&  8.847E$-$04&  0.728 \\
 19&   4& $-$& $4f^{11}~(^4I^1)~5d~^5G$ &      9&   4& +& $4f^{12}~^1G$            &     7875&    12697.85&  9.995E+01&  2.174E$-$05&  0.682 \\
 20&   5& $-$& $4f^{11}~(^4I^1)~5d~^5H$ &      1&   6& +& $4f^{12}~^3H$            &    26463&     3778.78&  2.335E+05&  5.499E$-$03&  0.746 \\
 20&   5& $-$& $4f^{11}~(^4I^1)~5d~^5H$ &      2&   4& +& $4f^{12}~^3F$            &    20732&     4823.32&  3.069E+05&  1.177E$-$02&  0.615 \\
 20&   5& $-$& $4f^{11}~(^4I^1)~5d~^5H$ &      3&   5& +& $4f^{12}~^3H$            &    19649&     5089.10&  3.843E+01&  1.641E$-$06&  0.983 \\
 20&   5& $-$& $4f^{11}~(^4I^1)~5d~^5H$ &      4&   4& +& $4f^{12}~^3H$            &    15573&     6420.98&  6.119E+04&  4.160E$-$03&  0.701 \\
\enddata
\tablecomments{All transition data are in length form. $dT$ is the relative difference of the transition rates in length and velocity form as given by equation \ref{accuracy}. Table~\ref{transition_data} is published in its entirety in the machine-readable format. Part of the values are shown here for guidance regarding its form and content.}
\end{deluxetable*}

\section{Summary and conclusion}
In the present paper energy levels of the ground [Xe]$4f^{12}$ and first excited [Xe]$4f^{11}5d$ configurations
for Er$^{2+}$ ion were computed using the GRASP2018 package. 
Transition data for E1 transitions between computed states are presented.
The accuracy of the obtained results is evaluated.

From the studies of the Er$^{2+}$ ion, and also from the previous investigations of Nd ions, 
it was observed that in such calculations to get the correct order of ground and excited configurations it is important 
to freeze the wave functions of ground configuration.

The valence-valence, core-valence, and core-core electron correlations were studied using different strategies.
This analysis has led to the final results in which the main balance electron correlation effects (mainly from VV substitutions) were included. This allows us to improve accuracy of the energy difference between different configurations considering the computational resources needed for the computations of such a complex system.

The rms deviations of the final results (using \textbf{SD 5d ZF$^{MCDHF}$} strategy) 
from the NIST or SE data for states of the ground and excited configurations
are 649 cm$^{-1}$, and 747 cm$^{-1}$, respectively.

Having analyzed convergence trends and dependencies of the gauge parameter $G$, we propose, for the Er$^{2+}$ ion, to use transition rates in the Babushkin gauge.

There is a lack of atomic data for the lanthanides. The present study is a first step towards the goal 
to provide this data with an accuracy high 
enough for opacity modeling.

\acknowledgments

This research was funded by a grant (No. S-LJB-18-1) from the Research Council of Lithuania.
Computations presented in this paper were performed
at the High Performance Computing Center ``HPC Sauletekis'' of the
Faculty of Physics at Vilnius University. DK is grateful to the support by NINS program of Promoting Research by
Networking among Institutions (Grant Number 01411702).

\bibliography{reference}



\end{document}